\documentclass[12pt, twoside]{article}
\usepackage{latexsym}
\usepackage{amsmath, amsfonts, amscd, amssymb}
\usepackage{amsthm}
\usepackage{fancybox}
\usepackage{eufrak, euscript, oldgerm, graphics}
\usepackage{pst-all}        
\usepackage{pst-poly}     
\usepackage{multido}      
\newcommand{\aconn}{\mathcal{A}}

\newcommand{\aut}{{\mathcal{A}}ut}

\newcommand{\conn}{\mathcal{D}}

\newcommand{\curv}{R}

\newcommand{\field}{\mathfrak{F}}

\newcommand{\ricci}{\EuScript{R}}

\newcommand{\gelsp}{\mathfrak{M}}

\newcommand{\eh}{\mathfrak{E}\mathfrak{H}}

\newcommand{\gauge}{\mathcal{U}}

\newcommand{\gl}{\mathcal{G}\mathcal{L}}

\newcommand{\hil}{{\mathcal{H}}}

\newcommand{\Hom}{{\mathcal{H}}om}

\newcommand{\lsh}{{\mathcal{L}}}

\newcommand{\modl}{\mathbf{\mathcal{E}}}

\newcommand{\natf}{\EuScript{N}}

\newcommand{\omg}{\Omega}
\newcommand{\Omg}{\mathbf{\Omega}}

\newcommand{\proj}{{\mathfrak{P}}}

\newcommand{\cont}{\mathcal{C}^{0}}
\newcommand{\smooth}{\mathcal{C}^{\infty}}
\newcommand{\sconn}{\textsf{A}}

\newcommand{\stre}{{\mathcal{F}}}
\newcommand{\struc}{\mathbf{A}}

\newcommand{\bull}{{\scriptstyle\bullet}}
\newcommand{\ym}{\mathfrak{Y}\mathfrak{M}}
\newcommand{\unc}{\mathfrak{U}}
\newcommand{\com}{\mathbb{C}}

\newcommand{\R}{\mathbb{R}}

\makeatletter

\def\diagram{\m@th\leftwidth=\z@ \rightwidth=\z@ \topheight=\z@
\botheight=\z@ \setbox\@picbox\hbox\bgroup}

\def\enddiagram{\egroup\wd\@picbox\rightwidth\unitlength
\ht\@picbox\topheight\unitlength \dp\@picbox\botheight\unitlength
\hskip\leftwidth\unitlength\box\@picbox}

\def\bfig{\begin{diagram}}
\def\efig{\end{diagram}}
\newcount\wideness \newcount\leftwidth \newcount\rightwidth
\newcount\highness \newcount\topheight \newcount\botheight

\def\ratchet#1#2{\ifnum#1<#2 \global #1=#2 \fi}

\def\putbox(#1,#2)#3{%
\horsize{\wideness}{#3} \divide\wideness by 2 {\advance\wideness
by #1 \ratchet{\rightwidth}{\wideness}} {\advance\wideness by -#1
\ratchet{\leftwidth}{\wideness}} \vertsize{\highness}{#3}
\divide\highness by 2 {\advance\highness by #2
\ratchet{\topheight}{\highness}} {\advance\highness by -#2
\ratchet{\botheight}{\highness}} \put(#1,#2){\makebox(0,0){$#3$}}}

\def\putlbox(#1,#2)#3{%
\horsize{\wideness}{#3} {\advance\wideness by #1
\ratchet{\rightwidth}{\wideness}} {\ratchet{\leftwidth}{-#1}}
\vertsize{\highness}{#3} \divide\highness by 2 {\advance\highness
by #2 \ratchet{\topheight}{\highness}} {\advance\highness by -#2
\ratchet{\botheight}{\highness}}
\put(#1,#2){\makebox(0,0)[l]{$#3$}}}

\def\putrbox(#1,#2)#3{%
\horsize{\wideness}{#3} {\ratchet{\rightwidth}{#1}}
{\advance\wideness by -#1 \ratchet{\leftwidth}{\wideness}}
\vertsize{\highness}{#3} \divide\highness by 2 {\advance\highness
by #2 \ratchet{\topheight}{\highness}} {\advance\highness by -#2
\ratchet{\botheight}{\highness}}
\put(#1,#2){\makebox(0,0)[r]{$#3$}}}

\def\adjust[#1]{} 

\newcount \coefa
\newcount \coefb
\newcount \coefc
\newcount\tempcounta
\newcount\tempcountb
\newcount\tempcountc
\newcount\tempcountd
\newcount\xext
\newcount\yext
\newcount\xoff
\newcount\yoff
\newcount\gap%
\newcount\arrowtypea
\newcount\arrowtypeb
\newcount\arrowtypec
\newcount\arrowtyped
\newcount\arrowtypee
\newcount\height
\newcount\width
\newcount\xpos
\newcount\ypos
\newcount\run
\newcount\rise
\newcount\arrowlength
\newcount\halflength
\newcount\arrowtype
\newdimen\tempdimen
\newdimen\xlen
\newdimen\ylen
\newsavebox{\tempboxa}%
\newsavebox{\tempboxb}%
\newsavebox{\tempboxc}%

\newdimen\w@dth

\def\setw@dth#1#2{\setbox\z@\hbox{\m@th$#1$}\w@dth=\wd\z@
\setbox\@ne\hbox{\m@th$#2$}\ifnum\w@dth<\wd\@ne \w@dth=\wd\@ne \fi
\advance\w@dth by 1.2em}

\def\t@^#1_#2{\allowbreak\def\n@one{#1}\def\n@two{#2}\mathrel
{\setw@dth{#1}{#2} \mathop{\hbox to
\w@dth{\rightarrowfill}}\limits \ifx\n@one\empty\else
^{\box\z@}\fi \ifx\n@two\empty\else _{\box\@ne}\fi}}
\def\t@@^#1{\@ifnextchar_{\t@^{#1}}{\t@^{#1}_{}}}
\def\to{\@ifnextchar^{\t@@}{\t@@^{}}}

\def\t@left^#1_#2{\def\n@one{#1}\def\n@two{#2}\mathrel{\setw@dth{#1}{#2}
\mathop{\hbox to \w@dth{\leftarrowfill}}\limits
\ifx\n@one\empty\else ^{\box\z@}\fi \ifx\n@two\empty\else
_{\box\@ne}\fi}}
\def\t@@left^#1{\@ifnextchar_{\t@left^{#1}}{\t@left^{#1}_{}}}
\def\toleft{\@ifnextchar^{\t@@left}{\t@@left^{}}}

\def\two@^#1_#2{\allowbreak
\def\n@one{#1}\def\n@two{#2}\mathrel{\setw@dth{#1}{#2}
\mathop{\vcenter{\lineskip\z@\baselineskip\z@
                 \hbox to \w@dth{\rightarrowfill}%
                 \hbox to \w@dth{\rightarrowfill}}%
       }\limits
\ifx\n@one\empty\else ^{\box\z@}\fi \ifx\n@two\empty\else
_{\box\@ne}\fi}}
\def\tw@@^#1{\@ifnextchar _{\two@^{#1}}{\two@^{#1}_{}}}
\def\two{\@ifnextchar ^{\tw@@}{\tw@@^{}}}

\def\tofr@^#1_#2{\def\n@one{#1}\def\n@two{#2}\mathrel{\setw@dth{#1}{#2}
\mathop{\vcenter{\hbox to \w@dth{\rightarrowfill}\kern-1.7ex
                 \hbox to \w@dth{\leftarrowfill}}%
       }\limits
\ifx\n@one\empty\else ^{\box\z@}\fi \ifx\n@two\empty\else
_{\box\@ne}\fi}}
\def\t@fr@^#1{\@ifnextchar_ {\tofr@^{#1}}{\tofr@^{#1}_{}}}
\def\tofro{\@ifnextchar^ {\t@fr@}{\t@fr@^{}}}

\def\mon{\mathop{\m@th\hbox to
      14.6\P@{\lasyb\char'51\hskip-2.1\P@$\arrext$\hss
$\mathord\rightarrow$}}\limits} 
\def\leftmono{\mathrel{\m@th\hbox to
14.6\P@{$\mathord\leftarrow$\hss$\arrext$\hskip-2.1\P@\lasyb\char'50%
}}\limits} 
\mathchardef\arrext="0200       

\def\settypes(#1,#2,#3){\arrowtypea#1 \arrowtypeb#2 \arrowtypec#3}
\def\settoheight#1#2{\setbox\@tempboxa\hbox{#2}#1\ht\@tempboxa\relax}%
\def\settodepth#1#2{\setbox\@tempboxa\hbox{#2}#1\dp\@tempboxa\relax}%
\def\settokens`#1`#2`#3`#4`{%
     \def\tokena{#1}\def\tokenb{#2}\def\tokenc{#3}\def\tokend{#4}}
\def\setsqparms[#1`#2`#3`#4;#5`#6]{%
\arrowtypea #1 \arrowtypeb #2 \arrowtypec #3 \arrowtyped #4 \width
#5 \height #6 }
\def\setpos(#1,#2){\xpos=#1 \ypos#2}

\def\settriparms[#1`#2`#3;#4]{\settripairparms[#1`#2`#3`1`1;#4]}%

\def\settripairparms[#1`#2`#3`#4`#5;#6]{%
\arrowtypea #1 \arrowtypeb #2 \arrowtypec #3 \arrowtyped #4
\arrowtypee #5 \width #6 \height #6 }

\def\resetparms{\settripairparms[1`1`1`1`1;500]\width 500}

\resetparms

\def\mvector(#1,#2)#3{
\put(0,0){\vector(#1,#2){#3}}%
\put(0,0){\vector(#1,#2){26}}%
}
\def\evector(#1,#2)#3{{
\arrowlength #3
\put(0,0){\vector(#1,#2){\arrowlength}}%
\advance \arrowlength by-30
\put(0,0){\vector(#1,#2){\arrowlength}}%
}}

\def\horsize#1#2{%
\settowidth{\tempdimen}{$#2$}%
#1=\tempdimen \divide #1 by\unitlength }

\def\vertsize#1#2{%
\settoheight{\tempdimen}{$#2$}%
#1=\tempdimen
\settodepth{\tempdimen}{$#2$}%
\advance #1 by\tempdimen \divide #1 by\unitlength }

\def\putvector(#1,#2)(#3,#4)#5#6{{%
\ifnum3<\arrowtype \putdashvector(#1,#2)(#3,#4)#5\arrowtype \else
\ifnum\arrowtype<-3 \putdashvector(#1,#2)(#3,#4)#5\arrowtype \else
\xpos=#1 \ypos=#2 \run=#3 \rise=#4 \arrowlength=#5 \ifnum
\arrowtype<0
    \ifnum \run=0
        \advance \ypos by-\arrowlength
    \else
        \tempcounta \arrowlength
        \multiply \tempcounta by\rise
        \divide \tempcounta by\run
        \ifnum\run>0
            \advance \xpos by\arrowlength
            \advance \ypos by\tempcounta
        \else
            \advance \xpos by-\arrowlength
            \advance \ypos by-\tempcounta
        \fi
    \fi
    \multiply \arrowtype by-1
    \multiply \rise by-1
    \multiply \run by-1
\fi \ifcase \arrowtype
\or \put(\xpos,\ypos){\vector(\run,\rise){\arrowlength}}%
\or \put(\xpos,\ypos){\mvector(\run,\rise)\arrowlength}%
\or \put(\xpos,\ypos){\evector(\run,\rise){\arrowlength}}%
\fi\fi\fi }}

\def\putsplitvector(#1,#2)#3#4{
\xpos #1 \ypos #2 \arrowtype #4 \halflength #3 \arrowlength #3
\gap 140 \advance \halflength by-\gap \divide \halflength by2
\ifnum\arrowtype>0
   \ifcase \arrowtype
   \or \put(\xpos,\ypos){\line(0,-1){\halflength}}%
       \advance\ypos by-\halflength
       \advance\ypos by-\gap
       \put(\xpos,\ypos){\vector(0,-1){\halflength}}%
   \or \put(\xpos,\ypos){\line(0,-1)\halflength}%
       \put(\xpos,\ypos){\vector(0,-1)3}%
       \advance\ypos by-\halflength
       \advance\ypos by-\gap
       \put(\xpos,\ypos){\vector(0,-1){\halflength}}%
   \or \put(\xpos,\ypos){\line(0,-1)\halflength}%
       \advance\ypos by-\halflength
       \advance\ypos by-\gap
       \put(\xpos,\ypos){\evector(0,-1){\halflength}}%
   \fi
\else \arrowtype=-\arrowtype
   \ifcase\arrowtype
   \or \advance \ypos by-\arrowlength
       \put(\xpos,\ypos){\line(0,1){\halflength}}%
       \advance\ypos by\halflength
       \advance\ypos by\gap
       \put(\xpos,\ypos){\vector(0,1){\halflength}}%
   \or \advance \ypos by-\arrowlength
       \put(\xpos,\ypos){\line(0,1)\halflength}%
       \put(\xpos,\ypos){\vector(0,1)3}%
       \advance\ypos by\halflength
       \advance\ypos by\gap
       \put(\xpos,\ypos){\vector(0,1){\halflength}}%
   \or \advance \ypos by-\arrowlength
       \put(\xpos,\ypos){\line(0,1)\halflength}%
       \advance\ypos by\halflength
       \advance\ypos by\gap
       \put(\xpos,\ypos){\evector(0,1){\halflength}}%
   \fi
\fi }

\def\putmorphism(#1)(#2,#3)[#4`#5`#6]#7#8#9{{%
\run #2 \rise #3 \ifnum\rise=0
  \puthmorphism(#1)[#4`#5`#6]{#7}{#8}#9%
\else\ifnum\run=0
  \putvmorphism(#1)[#4`#5`#6]{#7}{#8}#9%
\else
\setpos(#1)%
\arrowlength #7 \arrowtype #8 \ifnum\run=0 \else\ifnum\rise=0
\else \ifnum\run>0
    \coefa=1
\else
   \coefa=-1
\fi \ifnum\arrowtype>0
   \coefb=0
   \coefc=-1
\else
   \coefb=\coefa
   \coefc=1
   \arrowtype=-\arrowtype
\fi \width=2 \multiply \width by\run \divide \width by\rise \ifnum
\width<0  \width=-\width\fi \advance\width by60 \if l#9
\width=-\width\fi
\putbox(\xpos,\ypos){#4}
{\multiply \coefa by\arrowlength
\advance\xpos by\coefa \multiply \coefa by\rise \divide \coefa
by\run \advance \ypos by\coefa
\putbox(\xpos,\ypos){#5} }%
{\multiply \coefa by\arrowlength
\divide \coefa by2 \advance \xpos by\coefa \advance \xpos by\width
\multiply \coefa by\rise \divide \coefa by\run \advance \ypos
by\coefa
\if l#9%
   \putrbox(\xpos,\ypos){#6}%
\else\if r#9%
   \putlbox(\xpos,\ypos){#6}%
\fi\fi }%
{\multiply \rise by-\coefc
\multiply \run by-\coefc \multiply \coefb by\arrowlength \advance
\xpos by\coefb \multiply \coefb by\rise \divide \coefb by\run
\advance \ypos by\coefb \multiply \coefc by70 \advance \ypos
by\coefc \multiply \coefc by\run \divide \coefc by\rise \advance
\xpos by\coefc \multiply \coefa by140 \multiply \coefa by\run
\divide \coefa by\rise \advance \arrowlength by\coefa
\ifcase\arrowtype
\or \put(\xpos,\ypos){\vector(\run,\rise){\arrowlength}}%
\or \put(\xpos,\ypos){\mvector(\run,\rise){\arrowlength}}%
\or \put(\xpos,\ypos){\evector(\run,\rise){\arrowlength}}%
\fi}\fi\fi\fi\fi}}

\newcount\numbdashes \newcount\lengthdash \newcount\increment

\def\howmanydashes{
\numbdashes=\arrowlength \lengthdash=40 \divide\numbdashes by
\lengthdash \lengthdash=\arrowlength \divide\lengthdash by
\numbdashes
\increment=\lengthdash \multiply\lengthdash by 3
\divide\lengthdash by 5 }

\def\putdashvector(#1)(#2,#3)#4#5{%
\ifnum#3=0 \putdashhvector(#1){#4}#5 \else \ifnum#2=0
\putdashvvector(#1){#4}#5\fi\fi}

\def\putdashhvector(#1,#2)#3#4{{%
\arrowlength=#3 \howmanydashes
\multiput(#1,#2)(\increment,0){\numbdashes}%
{\vrule height .4pt width \lengthdash\unitlength} \arrowtype=#4
\xpos=#1 \ifnum\arrowtype<0 \advance\arrowtype by 7 \fi
\ifcase\arrowtype \or \advance\xpos by 10
    \put(\xpos,#2){\vector(-1,0){\lengthdash}}
    \advance\xpos by 40
    \put(\xpos,#2){\vector(-1,0){\lengthdash}}
\or \advance \xpos by 10
    \put(\xpos,#2){\vector(-1,0){\lengthdash}}
    \advance\xpos by  \arrowlength
    \advance\xpos by  -50
    \put(\xpos,#2){\vector(-1,0){\lengthdash}}
\or \advance\xpos by 10
    \put(\xpos,#2){\vector(-1,0){\lengthdash}}
\or \advance\xpos by \arrowlength
    \advance\xpos by -\lengthdash
    \put(\xpos,#2){\vector(1,0){\lengthdash}}
\or {\advance\xpos by 10
    \put(\xpos,#2){\vector(1,0){\lengthdash}}}
    \advance\xpos by \arrowlength
    \advance\xpos by -\lengthdash
    \put(\xpos,#2){\vector(1,0){\lengthdash}}
\or \advance\xpos by \arrowlength
    \advance\xpos by -\lengthdash
    \put(\xpos,#2){\vector(1,0){\lengthdash}}
    \advance\xpos by -40
    \put(\xpos,#2){\vector(1,0){\lengthdash}}
   \fi
}}

\def\putdashvvector(#1,#2)#3#4{{%
\arrowlength=#3 \howmanydashes \ypos=#2 \advance\ypos by
-\arrowlength
\multiput(#1,#2)(0,\increment){\numbdashes}%
    {\vrule width .4pt height \lengthdash\unitlength}
\arrowtype=#4 \ypos=#2 \ifnum\arrowtype<0 \advance\arrowtype by 7
\fi \ifcase\arrowtype \or \advance\ypos by \arrowlength
\advance\ypos by -40
    \put(#1,\ypos){\vector(0,1){\lengthdash}}
    \advance\ypos by -40
    \put(#1,\ypos){\vector(0,1){\lengthdash}}
\or \advance\ypos by 10
    \put(#1,\ypos){\vector(0,1){\lengthdash}}
    \advance\ypos by \arrowlength \advance\ypos by -40
    \put(#1,\ypos){\vector(0,1){\lengthdash}}
\or \advance\ypos by \arrowlength \advance\ypos by -40
    \put(#1,\ypos){\vector(0,1){\lengthdash}}
\or \advance\ypos by 10
    \put(#1,\ypos){\vector(0,-1){\lengthdash}}
\or \advance\ypos by 10
    \put(#1,\ypos){\vector(0,-1){\lengthdash}}
    \advance\ypos by \arrowlength \advance\ypos by -40
    \put(#1,\ypos){\vector(0,-1){\lengthdash}}
\or \advance\ypos by 10
    \put(#1,\ypos){\vector(0,-1){\lengthdash}}
    \advance\ypos by 40
    \put(#1,\ypos){\vector(0,-1){\lengthdash}}
\fi }}

\def\puthmorphism(#1,#2)[#3`#4`#5]#6#7#8{{%
\xpos #1 \ypos #2 \width #6 \arrowlength #6 \arrowtype=#7
\putbox(\xpos,\ypos){#3\vphantom{#4}}%
{\advance \xpos by\arrowlength
\putbox(\xpos,\ypos){\vphantom{#3}#4}}%
\horsize{\tempcounta}{#3}%
\horsize{\tempcountb}{#4}%
\divide \tempcounta by2 \divide \tempcountb by2 \advance
\tempcounta by30 \advance \tempcountb by30 \advance \xpos
by\tempcounta \advance \arrowlength by-\tempcounta \advance
\arrowlength by-\tempcountb
\putvector(\xpos,\ypos)(1,0)\arrowlength\arrowtype \divide
\arrowlength by2 \advance \xpos by\arrowlength
\vertsize{\tempcounta}{#5}%
\divide\tempcounta by2 \advance \tempcounta by20
\if a#8 %
   \advance \ypos by\tempcounta
   \putbox(\xpos,\ypos){#5}%
\else
   \advance \ypos by-\tempcounta
   \putbox(\xpos,\ypos){#5}%
\fi}}

\def\putvmorphism(#1,#2)[#3`#4`#5]#6#7#8{{%
\xpos #1 \ypos #2 \arrowlength #6 \arrowtype #7
\settowidth{\xlen}{$#5$}%
\putbox(\xpos,\ypos){#3}%
{\advance \ypos by-\arrowlength
\putbox(\xpos,\ypos){#4}}%
{\advance\arrowlength by-140 \advance \ypos by-70 \ifdim\xlen>0pt
   \if m#8%
      \putsplitvector(\xpos,\ypos)\arrowlength\arrowtype
   \else
   \putvector(\xpos,\ypos)(0,-1)\arrowlength\arrowtype
   \fi
\else
   \putvector(\xpos,\ypos)(0,-1)\arrowlength\arrowtype
\fi}%
\ifdim\xlen>0pt
   \divide \arrowlength by2
   \advance\ypos by-\arrowlength
   \if l#8%
      \advance \xpos by-40
      \putrbox(\xpos,\ypos){#5}%
   \else\if r#8%
      \advance \xpos by40
      \putlbox(\xpos,\ypos){#5}%
   \else
      \putbox(\xpos,\ypos){#5}%
   \fi\fi
\fi }}

\def\putsquarep<#1>(#2)[#3;#4`#5`#6`#7]{{%
\setsqparms[#1]%
\setpos(#2)%
\settokens`#3`%
\puthmorphism(\xpos,\ypos)[\tokenc`\tokend`{#7}]{\width}{\arrowtyped}b%
\advance\ypos by \height
\puthmorphism(\xpos,\ypos)[\tokena`\tokenb`{#4}]{\width}{\arrowtypea}a%
\putvmorphism(\xpos,\ypos)[``{#5}]{\height}{\arrowtypeb}l%
\advance\xpos by \width
\putvmorphism(\xpos,\ypos)[``{#6}]{\height}{\arrowtypec}r%
}}

\def\putsquare{\@ifnextchar <{\putsquarep}{\putsquarep%
   <\arrowtypea`\arrowtypeb`\arrowtypec`\arrowtyped;\width`\height>}}
\def\square{\@ifnextchar< {\squarep}{\squarep
   <\arrowtypea`\arrowtypeb`\arrowtypec`\arrowtyped;\width`\height>}}
\def\squarep<#1>[#2`#3`#4`#5;#6`#7`#8`#9]{{
\setsqparms[#1]
\diagram
\putsquarep<\arrowtypea`\arrowtypeb`\arrowtypec`
\arrowtyped;\width`\height>
(0,0)[#2`#3`#4`{#5};#6`#7`#8`{#9}]
\enddiagram
}}                                                 
\def\putptrianglep<#1>(#2,#3)[#4`#5`#6;#7`#8`#9]{{%
\settriparms[#1]%
\xpos=#2 \ypos=#3 \advance\ypos by \height
\puthmorphism(\xpos,\ypos)[#4`#5`{#7}]{\height}{\arrowtypea}a%
\putvmorphism(\xpos,\ypos)[`#6`{#8}]{\height}{\arrowtypeb}l%
\advance\xpos by\height
\putmorphism(\xpos,\ypos)(-1,-1)[``{#9}]{\height}{\arrowtypec}r%
}}

\def\putptriangle{\@ifnextchar <{\putptrianglep}{\putptrianglep
   <\arrowtypea`\arrowtypeb`\arrowtypec;\height>}}
\def\ptriangle{\@ifnextchar <{\ptrianglep}{\ptrianglep
   <\arrowtypea`\arrowtypeb`\arrowtypec;\height>}}
\def\ptrianglep<#1>[#2`#3`#4;#5`#6`#7]{{
\settriparms[#1]
\diagram
\putptrianglep<\arrowtypea`\arrowtypeb`
\arrowtypec;\height>
(0,0)[#2`#3`#4;#5`#6`{#7}]
\enddiagram
}}                                            

\def\putqtrianglep<#1>(#2,#3)[#4`#5`#6;#7`#8`#9]{{%
\settriparms[#1]%
\xpos=#2 \ypos=#3 \advance\ypos by\height
\puthmorphism(\xpos,\ypos)[#4`#5`{#7}]{\height}{\arrowtypea}a%
\putmorphism(\xpos,\ypos)(1,-1)[``{#8}]{\height}{\arrowtypeb}l%
\advance\xpos by\height
\putvmorphism(\xpos,\ypos)[`#6`{#9}]{\height}{\arrowtypec}r%
}}

\def\putqtriangle{\@ifnextchar <{\putqtrianglep}{\putqtrianglep
   <\arrowtypea`\arrowtypeb`\arrowtypec;\height>}}
\def\qtriangle{\@ifnextchar <{\qtrianglep}{\qtrianglep
   <\arrowtypea`\arrowtypeb`\arrowtypec;\height>}}
\def\qtrianglep<#1>[#2`#3`#4;#5`#6`#7]{{
\settriparms[#1]
\width=\height                                
\diagram
\putqtrianglep<\arrowtypea`\arrowtypeb`
\arrowtypec;\height>
(0,0)[#2`#3`#4;#5`#6`{#7}]
\enddiagram
}}

\def\putdtrianglep<#1>(#2,#3)[#4`#5`#6;#7`#8`#9]{{%
\settriparms[#1]%
\xpos=#2 \ypos=#3
\puthmorphism(\xpos,\ypos)[#5`#6`{#9}]{\height}{\arrowtypec}b%
\advance\xpos by \height \advance\ypos by\height
\putmorphism(\xpos,\ypos)(-1,-1)[``{#7}]{\height}{\arrowtypea}l%
\putvmorphism(\xpos,\ypos)[#4``{#8}]{\height}{\arrowtypeb}r%
}}

\def\putdtriangle{\@ifnextchar <{\putdtrianglep}{\putdtrianglep
   <\arrowtypea`\arrowtypeb`\arrowtypec;\height>}}
\def\dtriangle{\@ifnextchar <{\dtrianglep}{\dtrianglep
   <\arrowtypea`\arrowtypeb`\arrowtypec;\height>}}
\def\dtrianglep<#1>[#2`#3`#4;#5`#6`#7]{{
\settriparms[#1]
\width=\height                                
\diagram
\putdtrianglep<\arrowtypea`\arrowtypeb`
\arrowtypec;\height>
(0,0)[#2`#3`#4;#5`#6`{#7}]
\enddiagram
}}

\def\putbtrianglep<#1>(#2,#3)[#4`#5`#6;#7`#8`#9]{{%
\settriparms[#1]%
\xpos=#2 \ypos=#3
\puthmorphism(\xpos,\ypos)[#5`#6`{#9}]{\height}{\arrowtypec}b%
\advance\ypos by\height
\putmorphism(\xpos,\ypos)(1,-1)[``{#8}]{\height}{\arrowtypeb}r%
\putvmorphism(\xpos,\ypos)[#4``{#7}]{\height}{\arrowtypea}l%
}}

\def\putbtriangle{\@ifnextchar <{\putbtrianglep}{\putbtrianglep
   <\arrowtypea`\arrowtypeb`\arrowtypec;\height>}}
\def\btriangle{\@ifnextchar <{\btrianglep}{\btrianglep
   <\arrowtypea`\arrowtypeb`\arrowtypec;\height>}}
\def\btrianglep<#1>[#2`#3`#4;#5`#6`#7]{{
\settriparms[#1]
\width=\height                               
\diagram
\putbtrianglep<\arrowtypea`\arrowtypeb`
\arrowtypec;\height>
(0,0)[#2`#3`#4;#5`#6`{#7}]
\enddiagram
}}

\def\putAtrianglep<#1>(#2,#3)[#4`#5`#6;#7`#8`#9]{{%
\settriparms[#1]%
\xpos=#2 \ypos=#3 {\multiply \height by2
\puthmorphism(\xpos,\ypos)[#5`#6`{#9}]{\height}{\arrowtypec}b}%
\advance\xpos by\height \advance\ypos by\height
\putmorphism(\xpos,\ypos)(-1,-1)[#4``{#7}]{\height}{\arrowtypea}l%
\putmorphism(\xpos,\ypos)(1,-1)[``{#8}]{\height}{\arrowtypeb}r%
}}

\def\putAtriangle{\@ifnextchar <{\putAtrianglep}{\putAtrianglep
   <\arrowtypea`\arrowtypeb`\arrowtypec;\height>}}
\def\Atriangle{\@ifnextchar <{\Atrianglep}{\Atrianglep
   <\arrowtypea`\arrowtypeb`\arrowtypec;\height>}}
\def\Atrianglep<#1>[#2`#3`#4;#5`#6`#7]{{
\settriparms[#1]
\width=\height                                     
\diagram
\putAtrianglep<\arrowtypea`\arrowtypeb`
\arrowtypec;\height>
(0,0)[#2`#3`#4;#5`#6`{#7}]
\enddiagram
}}

\def\putAtrianglepairp<#1>(#2)[#3;#4`#5`#6`#7`#8]{{%
\settripairparms[#1]%
\setpos(#2)%
\settokens`#3`%
\puthmorphism(\xpos,\ypos)[\tokenb`\tokenc`{#7}]{\height}{\arrowtyped}b%
\advance\xpos by\height
\puthmorphism(\xpos,\ypos)[\phantom{\tokenc}`\tokend`{#8}]%
{\height}{\arrowtypee}b%
\advance\ypos by\height
\putmorphism(\xpos,\ypos)(-1,-1)[\tokena``{#4}]{\height}{\arrowtypea}l%
\putvmorphism(\xpos,\ypos)[``{#5}]{\height}{\arrowtypeb}m%
\putmorphism(\xpos,\ypos)(1,-1)[``{#6}]{\height}{\arrowtypec}r%
}}

\def\putAtrianglepair{\@ifnextchar <{\putAtrianglepairp}{\putAtrianglepairp%
   <\arrowtypea`\arrowtypeb`\arrowtypec`\arrowtyped`\arrowtypee;\height>}}
\def\Atrianglepair{\@ifnextchar <{\Atrianglepairp}{\Atrianglepairp%
   <\arrowtypea`\arrowtypeb`\arrowtypec`\arrowtyped`\arrowtypee;\height>}}

\def\Atrianglepairp<#1>[#2;#3`#4`#5`#6`#7]{{
\settripairparms[#1]
\settokens`#2`
\width=\height                                
\diagram
\putAtrianglepairp                            
<\arrowtypea`\arrowtypeb`\arrowtypec`
\arrowtyped`\arrowtypee;\height>
(0,0)[{#2};#3`#4`#5`#6`{#7}]
\enddiagram
}}

\def\putVtrianglep<#1>(#2,#3)[#4`#5`#6;#7`#8`#9]{{%
\settriparms[#1]%
\xpos=#2 \ypos=#3 \advance\ypos by\height {\multiply\height by2
\puthmorphism(\xpos,\ypos)[#4`#5`{#7}]{\height}{\arrowtypea}a}%
\putmorphism(\xpos,\ypos)(1,-1)[`#6`{#8}]{\height}{\arrowtypeb}l%
\advance\xpos by\height \advance\xpos by\height
\putmorphism(\xpos,\ypos)(-1,-1)[``{#9}]{\height}{\arrowtypec}r%
}}

\def\putVtriangle{\@ifnextchar <{\putVtrianglep}{\putVtrianglep
   <\arrowtypea`\arrowtypeb`\arrowtypec;\height>}}
\def\Vtriangle{\@ifnextchar <{\Vtrianglep}{\Vtrianglep
   <\arrowtypea`\arrowtypeb`\arrowtypec;\height>}}
\def\Vtrianglep<#1>[#2`#3`#4;#5`#6`#7]{{
\settriparms[#1]
\width=\height                                 
\diagram
\putVtrianglep<\arrowtypea`\arrowtypeb`
\arrowtypec;\height>
(0,0)[#2`#3`#4;#5`#6`{#7}]
\enddiagram
}}

\def\putVtrianglepairp<#1>(#2)[#3;#4`#5`#6`#7`#8]{{
\settripairparms[#1]%
\setpos(#2)%
\settokens`#3`%
\advance\ypos by\height
\putmorphism(\xpos,\ypos)(1,-1)[`\tokend`{#6}]{\height}{\arrowtypec}l%
\puthmorphism(\xpos,\ypos)[\tokena`\tokenb`{#4}]{\height}{\arrowtypea}a%
\advance\xpos by\height
\puthmorphism(\xpos,\ypos)[\phantom{\tokenb}`\tokenc`{#5}]%
{\height}{\arrowtypeb}a%
\putvmorphism(\xpos,\ypos)[``{#7}]{\height}{\arrowtyped}m%
\advance\xpos by\height
\putmorphism(\xpos,\ypos)(-1,-1)[``{#8}]{\height}{\arrowtypee}r%
}}

\def\putVtrianglepair{\@ifnextchar <{\putVtrianglepairp}{\putVtrianglepairp%
    <\arrowtypea`\arrowtypeb`\arrowtypec`\arrowtyped`\arrowtypee;\height>}}
\def\Vtrianglepair{\@ifnextchar <{\Vtrianglepairp}{\Vtrianglepairp%
    <\arrowtypea`\arrowtypeb`\arrowtypec`\arrowtyped`\arrowtypee;\height>}}
\def\Vtrianglepairp<#1>[#2;#3`#4`#5`#6`#7]{{
\settripairparms[#1]
\settokens`#2`
\diagram
\putVtrianglepairp                             
<\arrowtypea`\arrowtypeb`\arrowtypec`
\arrowtyped`\arrowtypee;\height>
(0,0)[{#2};#3`#4`#5`#6`{#7}]
\enddiagram
}}

\def\putCtrianglep<#1>(#2,#3)[#4`#5`#6;#7`#8`#9]{{%
\settriparms[#1]%
\xpos=#2 \ypos=#3 \advance\ypos by\height
\putmorphism(\xpos,\ypos)(1,-1)[``{#9}]{\height}{\arrowtypec}l%
\advance\xpos by\height \advance\ypos by\height
\putmorphism(\xpos,\ypos)(-1,-1)[#4`#5`{#7}]{\height}{\arrowtypea}l%
{\multiply\height by 2
\putvmorphism(\xpos,\ypos)[`#6`{#8}]{\height}{\arrowtypeb}r}%
}}

\def\putCtriangle{\@ifnextchar <{\putCtrianglep}{\putCtrianglep
    <\arrowtypea`\arrowtypeb`\arrowtypec;\height>}}
\def\Ctriangle{\@ifnextchar <{\Ctrianglep}{\Ctrianglep
    <\arrowtypea`\arrowtypeb`\arrowtypec;\height>}}
\def\Ctrianglep<#1>[#2`#3`#4;#5`#6`#7]{{
\settriparms[#1]
\width=\height                               
\diagram
\putCtrianglep<\arrowtypea`\arrowtypeb`
\arrowtypec;\height>
(0,0)[#2`#3`#4;#5`#6`{#7}]
\enddiagram
}}                                           
\def\putDtrianglep<#1>(#2,#3)[#4`#5`#6;#7`#8`#9]{{%
\settriparms[#1]%
\xpos=#2 \ypos=#3 \advance\xpos by\height \advance\ypos by\height
\putmorphism(\xpos,\ypos)(-1,-1)[``{#9}]{\height}{\arrowtypec}r%
\advance\xpos by-\height \advance\ypos by\height
\putmorphism(\xpos,\ypos)(1,-1)[`#5`{#8}]{\height}{\arrowtypeb}r%
{\multiply\height by 2
\putvmorphism(\xpos,\ypos)[#4`#6`{#7}]{\height}{\arrowtypea}l}%
}}

\def\putDtriangle{\@ifnextchar <{\putDtrianglep}{\putDtrianglep
    <\arrowtypea`\arrowtypeb`\arrowtypec;\height>}}
\def\Dtriangle{\@ifnextchar <{\Dtrianglep}{\Dtrianglep
   <\arrowtypea`\arrowtypeb`\arrowtypec;\height>}}
\def\Dtrianglep<#1>[#2`#3`#4;#5`#6`#7]{{
\settriparms[#1]
\width=\height                              
\diagram
\putDtrianglep<\arrowtypea`\arrowtypeb`
\arrowtypec;\height>
(0,0)[#2`#3`#4;#5`#6`{#7}]
\enddiagram
}}                                          
\def\setrecparms[#1`#2]{\width=#1 \height=#2}%

\def\recursep<#1`#2>[#3;#4`#5`#6`#7`#8]{{\m@th
\width=#1 \height=#2 \settokens`#3`
\settowidth{\tempdimen}{$\tokena$} \ifdim\tempdimen=0pt
  \savebox{\tempboxa}{\hbox{$\tokenb$}}%
  \savebox{\tempboxb}{\hbox{$\tokend$}}%
  \savebox{\tempboxc}{\hbox{$#6$}}%
\else
  \savebox{\tempboxa}{\hbox{$\hbox{$\tokena$}\times\hbox{$\tokenb$}$}}%
  \savebox{\tempboxb}{\hbox{$\hbox{$\tokena$}\times\hbox{$\tokend$}$}}%
  \savebox{\tempboxc}{\hbox{$\hbox{$\tokena$}\times\hbox{$#6$}$}}%
\fi \ypos=\height \divide\ypos by 2 \xpos=\ypos \advance\xpos by
\width \bfig
\putCtrianglep<-1`1`1;\ypos>(0,0)[`\tokenc`;#5`#6`{#7}]%
\puthmorphism(\ypos,0)[\tokend`\usebox{\tempboxb}`{#8}]{\width}{-1}b%
\puthmorphism(\ypos,\height)[\tokenb`\usebox{\tempboxa}`{#4}]{\width}{-1}a%
\advance\ypos by \width
\putvmorphism(\ypos,\height)[``\usebox{\tempboxc}]{\height}1r%
\efig }}

\def\recurse{\@ifnextchar <{\recursep}{\recursep<\width`\height>}}

\def\puttwohmorphisms(#1,#2)[#3`#4;#5`#6]#7#8#9{{%
\puthmorphism(#1,#2)[#3`#4`]{#7}0a \ypos=#2 \advance\ypos by 20
\puthmorphism(#1,\ypos)[\phantom{#3}`\phantom{#4}`#5]{#7}{#8}a
\advance\ypos by -40
\puthmorphism(#1,\ypos)[\phantom{#3}`\phantom{#4}`#6]{#7}{#9}b }}

\def\puttwovmorphisms(#1,#2)[#3`#4;#5`#6]#7#8#9{{%
\putvmorphism(#1,#2)[#3`#4`]{#7}0a \xpos=#1 \advance\xpos by -20
\putvmorphism(\xpos,#2)[\phantom{#3}`\phantom{#4}`#5]{#7}{#8}l
\advance\xpos by 40
\putvmorphism(\xpos,#2)[\phantom{#3}`\phantom{#4}`#6]{#7}{#9}r }}

\def\puthcoequalizer(#1)[#2`#3`#4;#5`#6`#7]#8#9{{%
\setpos(#1)%
\puttwohmorphisms(\xpos,\ypos)[#2`#3;#5`#6]{#8}11%
\advance\xpos by #8
\puthmorphism(\xpos,\ypos)[\phantom{#3}`#4`#7]{#8}1{#9} }}

\def\putvcoequalizer(#1)[#2`#3`#4;#5`#6`#7]#8#9{{%
\setpos(#1)%
\puttwovmorphisms(\xpos,\ypos)[#2`#3;#5`#6]{#8}11%
\advance\ypos by -#8
\putvmorphism(\xpos,\ypos)[\phantom{#3}`#4`#7]{#8}1{#9} }}

\def\putthreehmorphisms(#1)[#2`#3;#4`#5`#6]#7(#8)#9{{%
\setpos(#1) \settypes(#8)
\if a#9 %
     \vertsize{\tempcounta}{#5}%
     \vertsize{\tempcountb}{#6}%
     \ifnum \tempcounta<\tempcountb \tempcounta=\tempcountb \fi
\else
     \vertsize{\tempcounta}{#4}%
     \vertsize{\tempcountb}{#5}%
     \ifnum \tempcounta<\tempcountb \tempcounta=\tempcountb \fi
\fi \advance \tempcounta by 60
\puthmorphism(\xpos,\ypos)[#2`#3`#5]{#7}{\arrowtypeb}{#9}
\advance\ypos by \tempcounta
\puthmorphism(\xpos,\ypos)[\phantom{#2}`\phantom{#3}`#4]{#7}{\arrowtypea}{#9}
\advance\ypos by -\tempcounta \advance\ypos by -\tempcounta
\puthmorphism(\xpos,\ypos)[\phantom{#2}`\phantom{#3}`#6]{#7}{\arrowtypec}{#9}
}}

\def\setarrowtoks[#1`#2`#3`#4`#5`#6]{%
\def\toka{#1}
\def\tokb{#2}
\def\tokc{#3}
\def\tokd{#4}
\def\toke{#5}
\def\tokf{#6}
}
\def\hex{\@ifnextchar <{\hexp}{\hexp<1000`400>}}
\def\hexp<#1`#2>[#3`#4`#5`#6`#7`#8;#9]{%
\setarrowtoks[#9] \yext=#2 \advance \yext by #2 \xext=#1
\advance\xext by \yext \bfig
\putCtriangle<-1`0`1;#2>(0,0)[`#5`;\tokb``\tokd] \xext=#1 \yext=#2
\advance \yext by #2
\putsquare<1`0`0`1;\xext`\yext>(#2,0)[#3`#4`#7`#8;\toka```\tokf]
\advance \xext by #2
\putDtriangle<0`1`-1;#2>(\xext,0)[`#6`;`\tokc`\toke] \efig }
\makeatother

\title{\bf\Large `Third' Quantization of Vacuum Einstein Gravity\\ and Free
Yang-Mills Theories\footnote{Wholeheartedly dedicated to Rafael
Sorkin on the occasion of his 60th birthday. Posted at the
e-archives: gr-qc/0606021.}}

\author{Ioannis Raptis\thanks{EU Marie Curie Reintegration Research Fellow, Algebra and Geometry Section,
Department of Mathematics, University of Athens,
Panepistimioupolis, Athens 157 84, Greece; {\em and} Visiting
Researcher, Theoretical Physics Group, Blackett Laboratory,
Imperial College of Science, Technology and Medicine, Prince
Consort Road, South Kensington, London SW7 2BZ, UK; e-mail:
i.raptis@ic.ac.uk}}

\date{All day today!}

\begin{document}

{\catcode`\ =13\global\let =\ \catcode`\^^M=13
\gdef^^M{\par\noindent}}
\def\verbatim{\tt
\catcode`\^^M=13 \catcode`\ =13 \catcode`\\=12 \catcode`\{=12
\catcode`\}=12 \catcode`\_=12 \catcode`\^=12 \catcode`\&=12
\catcode`\~=12 \catcode`\#=12 \catcode`\%=12 \catcode`\$=12
\catcode`|=0 }

\maketitle

\pagestyle{myheadings}\markboth{\centerline {\small {\sc {Ioannis
Raptis}}}}{\centerline {\footnotesize {\sc {`Third' Quantization
of Vacuum Einstein Gravity and Free Yang-Mills Theories}}}}

\pagenumbering{arabic}

\begin{abstract}

\noindent {\small Certain pivotal results from various
applications of Abstract Differential Geometry (ADG) to gravity
and gauge theories are presently collected and used to argue that
we already possess a geometrically (pre)quantized, second
quantized and manifestly background spacetime manifold independent
vacuum Einstein gravitational field dynamics. The arguments carry
also {\it mutatis mutandis} to the case of free Yang-Mills
theories, since from the ADG-theoretic perspective gravity is
regarded as another gauge field  theory. The powerful
algebraico-categorical, sheaf cohomological conceptual and
technical machinery of ADG is then employed, based on the
fundamental ADG-theoretic conception of a field as a pair $(\modl
,\conn)$ consisting of a vector sheaf $\modl$ and an algebraic
connection $\conn$ acting categorically as a sheaf morphism on
$\modl$'s  local sections, to introduce a `universal', because
expressly functorial, field quantization scenario coined {\em
third quantization}. Although third quantization is fully
covariant, on intuitive and heuristic grounds alone it formally
appears to follow a canonical route; albeit, in a purely algebraic
and, in contradistinction to geometric (pre)quantization and
(canonical) second quantization, manifestly background geometrical
spacetime manifold independent fashion, as befits ADG. All in all,
from the ADG-theoretic vantage, vacuum Einstein gravity and free
Yang-Mills theories are regarded as external spacetime manifold
unconstrained, third quantized, pure gauge field theories.  The
paper abounds with philosophical smatterings and speculative
remarks  about the potential import and significance of our
results to current and future Quantum Gravity research. A
postscript gives a brief account of this author's personal
encounters with Rafael Sorkin and his work.}

\vskip 0.1in



\end{abstract}

\setlength{\textwidth}{15.0cm} 
\setlength{\oddsidemargin}{0cm}  
\setlength{\evensidemargin}{0cm} 
\setlength{\topskip}{0pt}  
\setlength{\textheight}{21.0cm} 
\setlength{\footskip}{-2cm}

\setlength{\topmargin}{0pt}

\section{Motivational Remarks}

Modern fundamental physics may be cumulatively referred to as
`{\em field physics}'. The theoretical concept of `{\em field}' is
the cornerstone of our most successful and experimentally verified
theories of Nature: from the macroscopic General Relativity (GR)
describing gravity which shapes the large scale structure of the
Universe, to the microscopic Quantum Field Theory (QFT) describing
the structure and dynamical transmutations of matter at subatomic
scales \cite{auyang,cao}.

At the same time, field theory in general, at least as it has been
thought of and practiced almost ever since its inception until
today, appears to be inextricably tied to a background manifold,
which is physically interpreted as the `spacetime continuum'---be
it for example the curved Lorentzian spacetime manifold of GR, or
the flat Minkowski space of the flat (:gravity-free) QFTs of
matter. Indeed, the current theoretical consensus maintains that
it takes a mathematical continuum such as a locally Euclidean
space\footnote{Finite ({\it eg}, spacetime) or
infinite-dimensional manifolds ({\it eg}, the fields'
configuration spaces).} to accommodate systems with an infinite
number of degrees of freedom---the currently widely established
conception of fields. The bottom line is that field theory, at
least regarding the mathematical means that we have so far
employed to formulate it, relies heavily on the notions, methods
and technology of Classical Differential Geometry (CDG), which in
turn is vitally dependent on (the {\it a priori} assumption of) a
base differential manifold to support its concepts and
constructions \cite{gosch,kriele}. Let us reduce this to a `boxed
slogan':

\medskip

\centerline{\boxed{{\bf S1.}~{\rm
The~basic~mathematical~framework~of~field~theory~is~the~CDG~
of~smooth~manifolds.}}}

\medskip

\noindent In fact, such has been the influence of CDG on the
development of field theory (and vice versa!) that it is not an
exaggeration to say that {\em it is almost impossible to think of
the latter apart from the former}. One should consider for
instance the immense influence that the modern developments of CDG
in terms of smooth fiber bundles have exercised on the way we view
and treat (classical or quantum) gauge field theories of matter,
including gravity \cite{gosch,kriele,auyang,cao,ivanenko}.

Of course, that the theoretical physicist has so readily adopted
the mathematics of CDG may be largely attributed to the fact that
her principal aim---ideally, to discover and describe
(:mathematically model) the laws of Nature---coupled to her
theoretical requirement that the latter be {\em local}
mathematical expressions, have found fertile ground in the
manifold based CDG, as our second boxed slogan posits:

\medskip

\centerline{\boxed{{\bf S2.}~{\rm
Physical~laws~are~to~be~modelled~after~differential~equations}.}}

\medskip

\noindent For example, more than a century ago, Bertrand Russell
\cite{russell} went as far as to maintain that

\medskip

\centerline{``{\small\em The laws of physics can only be expressed
as differential equations.}''}

\medskip

\noindent Indeed, the background geometrical locally Euclidean
continuum (be it spacetime, or the field's configuration space)
provides one with a smooth geometrical platform on which the
apparently necessary {\em principle of infinitesimal
locality}---the {\it a priori} theoretical requirement for a
smooth causal nexus between the world-events triggered by
contiguous field actions---can be snugly accommodated and
(differential) geometrically pictured ({\it ie}, represented by
differential equations and their smooth solutions).

At the same time, few theoretical/mathematical
physicists---general relativists and quantum field theorists
alike---would disagree that the pointed background geometrical
spacetime manifold is the main culprit for various pathologies
that GR and QFT suffer from, such as singularities and related
unphysical infinities \cite{clarke4}. In principle, any point of
the underlying manifold can be the {\it locus} of a singularity of
some physically important smooth field---a site where the field
seems to blow up uncontrollably without bound and the law
(:differential equation) that it obeys appears to break down
somehow. Given that few physicists would actually admit such
divergences (:infinities) as being physical, it is remarkable that
even fewer would readily abandon the manifold based CDG as a
theoretical/mathematical framework in which to formulate and work
out field theories. They would rather resort to manifold and, {\it
in extenso}, CDG-conservative effective approximation ({\it eg},
perturbation) methods and would take great pains to devise quite
sophisticated regularization and renormalization techniques to
cope with the infinities, instead of doing away once and for all
with the background geometrical spacetime manifold $M$. In view of
S1, this is understandable, because if $M$ will have to go, so
will field theory, and then how, other than by differential
equations proper set up by CDG-means, would physical laws be
represented (S2)? {\it Mutatis mutandis} then for the geometrical
picturization of the local field and particle dynamics: how, other
than the usual imagery depicting the propagation and interaction
of fields (and their particles) on a continuous base spacetime
arena, could one geometrically picture (:represent) the field and
particle dynamics?\footnote{Think for example of the physicist's
`archetypical' theoretical image of {\em dynamical paths}
(:trajectories) traversed by particles during their dynamical
evolution. These are normally taken to be smooth curves in a
$4$-dimensional space-time continuum.}

Especially when it comes to Quantum Gravity (QG), the aforesaid
resort to CDG-conservative means may be justified on a reasonable
analogy (A) and its associated hopeful expectation (E); namely
that,

\begin{itemize}

\item {\bf A.} Much in the same way that the background manifold
conservative quantization of the classical CDG-based field
theories of matter managed to alleviate or even remove completely
the unphysical infinities via `analytic' renormalization ({\it
eg}, QED `resolving' the infinities of Maxwellian
electrodynamics), so QG---regarded as Quantum General Relativity
(QGR)\footnote{Here, by QGR we understand in general QG approached
as a QFT---a quantum (or quantized, canonically and/or
covariantly) field theory of gravity on a background differential
spacetime manifold \cite{thiem2,thiem3}, with the classical theory
being GR (on the same background!; see below).}---could (or more
demandingly, should!) remove singularities and their associated
infinities.

\item {\bf E.} Thus, all we have to aim and hope is for a better,
more subtle, refined and powerful `{\em
Analysis}'\footnote{Hereafter, the terms `CDG', `Analysis' and
`Differential Calculus on Manifolds' shall be regarded as synonyms
and used interchangeably.}---perhaps one with formal quantum
traits inherent in its formalism; albeit, one that still
essentially relies on a background geometrical manifold in one way
or another,\footnote{Here we have in mind various attempts at
applying a `quantized' ({\it eg}, `noncommutative') sort of
Calculus to quantum spacetime, gravity and gauge
theories---Connes' Noncommutative (Differential) Geometry being
the `canonical' example of such an enterprize
\cite{connes,connes1,kastler,madore,connes2,connes3}.} for {\em
how else could one do field theory}---be it {\em quantum} field
theory---{\em differential geometrically} (S1)?

\end{itemize}

\noindent Alas, QG has proven to be (perturbatively)
non-renormalizable, plus it appears to mandate the existence of a
fundamental space-time length-duration---the Planck scale---below
which the spacetime continuum is expected to give way to something
more reticular and quantal: `{\em quantized spacetime}', so to
speak. So, {\em there goes our cherished field-theoretic,
CDG-based, outlook on QG?} Not quite, yet.

{\it Prima facie}, in view of the existence of Planck's
fundamental cut-off spacetime scale, that QG (viewed and treated
as QGR) is non-renormalizable is not a blemish after all. Indeed,
the (perturbatively) {\em renormalizable} (flat) continuum based
(quantum) gauge theories describing the other three fundamental
forces do {\em not} have such dimensionful constants (:space-time
scales, or `coupling' constants combining to produce those scales)
inherent in their theoretical fabric. In turn, the Planck length
is the {\it raison d'\^{e}tre et de faire} of non-perturbative
QGR. The latter, at least in its present and most promising
gauge-theoretic formulation as Loop Quantum Gravity (LQG) and
Cosmology (LQC) \cite{rovsm1,thiem2,thiem3,smolin1} which is based
on the Ashtekar formalism for GR \cite{ash}, effectively removes
the continuous manifold picture of spacetime and resolves the
singularities-{\it cum}-infinities that the latter is responsible
for by means of `{\em spacetime quantization}'
\cite{rovsm,thiem1,ashlew3,ashlew4,boj1,modesto,husain}. However,
the mathematical formalism devised recently to formalize and carry
out that quantization, Quantum Riemannian Geometry (QRG)
\cite{ash4}, is still drawing amply from a background differential
manifold for its differential geometric expression. All in all,
the QRG-based LQG and LQC fulfill the aforementioned expectations
(A,E), and what's more, without resorting to perturbative
renormalization arguments, techniques or results. {\it En
passant}, let it be noted here that the other approach to
(non-perturbative) QG currently competing with LQG for popularity
(and monarchic hegemony!), (super)string theory (perturbative or
not), also heavily relies on the manifold based CDG for its
concepts and techniques. One should think for example of how
higher-dimensional (real analytic or holomorphic) differential
manifolds such as Riemann surfaces, K\"ahler spaces, Calabi-Yau
manifolds, $\mathbb{Z}_{2}$-graded manifolds (:supermanifolds),
{\it etc.}, have become the bread and butter mathematical
structures in current string and brane theory research.

With the remarks above in mind, an overarching theoretical
requirement or `principle' underlying most (if not all) of the
current (non-perturbative) QG approaches, including LQG and string
theory, is that of {\em background independence}
\cite{ash5,smolin2,seiberg}. Expressed as a theoretical
imperative:

\medskip

\centerline{\boxed{{\bf B.}~{\rm The~true~quantum~gravity~
must~be~a~background~independent~theory}.}}

\medskip

\noindent The original requirement for `background independence'
pertained to `background {\em geometry} indepe- ndence'---{\it
ie}, that QG should be formulated in a background {\em metric}
independent way. Lately, the term `geometry' is understood and
used in the broader (mathematical) sense of (a structureless set
endowed with some) `structure' \cite{maclane0}, so that a
background independent formulation of QG means one that does not
employ any fixed background structure---an `absolute geometrical
space' of any kind.\footnote{Hereafter, let us call a set equipped
with some structure a (mathematical) `{\em space}'. This is pretty
much how a `geometrical space' has been conceived in the physics
\cite{stachel2} and mathematics \cite{maclane0} literature.
Abiding by set-theoretic notions is not necessary, however. For
example, the novel `quantization on a category' scheme recently
proposed by Isham \cite{ish5,ish6,ish7,ish8}, may still be
perceived as being background dependent in the strict sense of B2
(see the next paragraph in the main text); albeit, the background
is not a point-set proper, but a category---a mathematical
universe of generalized (and in a certain sense, variable!) sets,
as well as of maps (:morphisms) between them. A die-hard
background independent `quantum gravitist' might still regard such
a scheme as being background dependent in disguise. But let us set
aside such `extremist' and `purist' views, and plough on. For in
any case, the backgrounds involved in Isham's work ({\it eg},
discrete topological spaces or causal sets) are far from being
smooth manifolds, and are by no means fixed.} These two
conceptions of background independence---the old, stricter and
`weaker' one, and the new, generalized and `stronger' one---are
pretty much how they have been recently expressed in \cite{kribs}
as a distillation from \cite{buttish0}:

\begin{quotation}
\noindent {\bf B1. `Weak' Background Independence (WBI):}
``{\small ...{\bf Background independence 1.} A quantum theory of
gravity is background independent if its basic quantities and
concepts do not presuppose a background metric.}''

\noindent {\bf B2. `Strong' Background independence (SBI):}
``{\small {\bf Background independence 2.} A quantum theory of
gravity is background independent if there is no fixed theoretical
structure. Any fixed structure will be regarded as a
background...}''
\end{quotation}

\noindent There are strong Leibnizian undertones in SBI, in the
following sense: the true QG must be formulated in a {\em
relational} way, without reference or recourse to any `absolute',
ether-like background structure ({\it eg}, `spacetime') whatsoever
\cite{smolin2}.

In this respect, it is fair to say that so far (non-perturbative)
string theory has {\em not} managed to achieve a background
independent formulation of QG even in the restricted (B1) sense,
since the whole formalism and interpretation of the theory vitally
depends on a background (usually taken to be Minkowski) metric
(space). Even LQG, although it is background metric independent
(WBI), it is not (yet) background independent in the stronger
sense (SBI), since, as noted earlier, its formulation relies
heavily on manifold based CDG-means---{\it ie}, the background in
this case being the geometrical differential (spacetime) manifold
and thus the theoretical/mathematical framework employed is
effectively the CDG of such smooth `domains'. The spacetime
continuum and its pathologies ({\it eg}, singularities and
associated infinities) is indeed evaded, but, as mentioned briefly
before, only after a canonical-type of quantization procedure is
exercised on the classical theory (:GR) and its supporting
spacetime continuum
\cite{rovsm,thiem1,ashlew3,ashlew4,boj1,modesto,husain}.

On the other hand, we have {\em Abstract Differential Geometry}
(ADG)---the purely algebraico-categorical (:sheaf-theoretic)
framework in which one can do differential geometry in a
manifestly background manifold independent, thus effectively
Calculus-free, way \cite{mall1,mall2,mall4}. Indeed, in a
Leibnizian-Machean sense \cite{malrap4}, the entire differential
geometric ADG-machinery focuses on, and derives directly from, the
algebraically ({\it ie}, sheaf theoretically) represented {\em
dynamical} relations between the `{\em geometrical objects}' that
`live' on `space(time)'---the dynamical fields
themselves---without that background `space(time)' playing {\em
any} role, thus having no physical significance whatsoever, in the
said field dynamics
\cite{mall1,mall2,malrap3,mall4,malrap1,malrap2,malrap3,malrap4}.
Moreover, the dynamics is still represented by differential
equations proper between the ADG-fields; albeit, the latter are
abstract, algebraico-categorical expressions involving equations
between sheaf morphisms that the fields are modelled after,
without recourse to a base spacetime manifold arena for their
geometrical support and interpretation (:`spacetime
picturization'). So far, ADG has enjoyed numerous applications to
gauge (:Yang-Mills) theories and gravity
\cite{mall1,mall2,mall3,mall4,mall9,malrap1,malrap2,
malrap3,malrap4,rap5,rap7,rap11}.

In the present paper we employ the purely algebraic
(:sheaf-theoretic) and manifestly background (spacetime)
manifoldless concepts and technology of ADG in order to arrive at
a `universal', because manifestly functorial, field-quantization
scenario for (free) Yang-Mills fields, including (vacuum) Einstein
gravity which from an ADG-theoretic perspective is regarded as
another gauge theory
\cite{mall3,malrap1,malrap2,malrap3,mall4,malrap4,rap5,rap7,rap11}.
The basic sheaf-theoretic machinery is {\em sheaf cohomology},
while the scenario formally resembles canonical quantization, but
it expressly avoids any mention of or reference to a background
geometrical (spacetime) manifold structure. Rather, it is an
entirely relational (:algebraic) scheme since, as befits ADG, it
concerns solely the ADG-fields (vacuum gravitational and free
Yang-Mills) involved. In particular, the said canonical
quntization-type of scenario involves our positing non-trivial
local commutation relations between certain characteristic local
(:differential) forms that uniquely characterize sheaf
cohomologically the ADG-gauge fields and their particle-quanta. In
turn, in a heuristic way these forms may be physically interpreted
as abstract position and momentum determinations (:`observables'),
hence the epithet `canonical' adjoined to the noun `quantization'
above. The base spacetime manifoldless sheaf cohomological
ADG-field quantization proposed here is coined `{\em third
quntization}' so as to distinguish it from the usual manifold and
CDG-based 2nd and, of course, 1st-quantization.

The paper is organized as follows: in the next section (2) we
recall certain pivotal results from various applications of ADG to
vacuum Einstein gravity (VEG) and free Yang-Mills (FYM) theories;
in particular, to the geometric (pre)quantization and second
quantization thereof. Based on these results, we then maintain
that we already possess a geometrically (pre)quantized and second
quantized vacuum Einstein gravitational and free Yang-Mills field
dynamics.  Especially, we highlight how the background spacetime
manifold independent ADG-formalism enables us to:

\begin{enumerate}

\item Extend the current so-called `gauge theory of the second
kind' (:{\em local} gauge field theory) to what is here coined
`{\em gauge field theory of the 3rd kind}'
\cite{malrap4,rap5,rap7,rap11}, which, although manifestly local
like its predecessor, it is not local(ized) on an external (to the
gauge fields themselves) geometrical spacetime continuum.

\item Effectively {\em halve the order of the formalism}, since in
our scheme the purely gauge connection field $\conn$, and not the
metric field $g_{\mu\nu}$ (or equivalently, the tetrad field
$e_{\mu}$), is the sole dynamical variable \cite{malrap3,
malrap4,rap5,rap7,rap11}. This enables us to contrast our purely
gauge-theoretic ADG-gravitational formalism against the {\em
manifestly background differential spacetime manifold dependent},
hence also CDG-based, second (:Einstein) and first order
(Ashtekar-Palatini) formalisms. Fittingly, we coin our approach
`{\em $\frac{1}{2}$-order formalism}'.

\item Pave the way towards {\em 3rd-quantization}, by evading
altogether a background spacetime manifold (thus also the
CDG-based 1st and 2nd-quantization scenaria), and by concentrating
instead on local algebraic dynamical relations between the
ADG-fields involved `in-themselves'.\footnote{This`{\em autonomy}'
of 3rd-quantization is its essential feature, and it makes us
think of the ADG-fields as quantum dynamically autonomous
(:self-supporting; physical laws' self-legislating) entities
reminiscent of Leibniz's `entelechian monads' \cite{leibniz}. This
analogy is consistent with the Leibnizian (:purely relational,
{\it ie}, algebraic) conception of the differential geometry
(:Differential Calculus) that ADG propounds (cf. \cite{malrap4}
and \cite{mall7,mall11,mall10} for an extensive discussion on
this). The autonomy of the dynamical ADG-fields will be further
corroborated by our 3rd-quantization scenario in the sequel.}

\end{enumerate}

\noindent Thus, in section 3 we entertain the possibility of
extending 2nd to 3rd-quantization in a way that suits the formal
ADG-gauge field theory of the 3rd kind and its physical semantics.
Third quantization is a heuristic conception of a dynamically
autonomous, because external (:background) spacetime manifoldless
and thereby unconstrained, ADG-gauge field quantization scenario
which is tailor-cut for the geometrically (pre)quantized and
second quantized ADG-field semantics. It formally appears to
follow a canonical route, since it involves non-trivial local
commutation relations between certain characteristic local
(:differential) forms that uniquely characterize sheaf
cohomologically the ADG-gauge fields and, from an ADG perspective
on 2nd and geometric (pre)quantization, their particle-quanta. In
turn, as noted above, in a heuristic way the said forms may be
physically interpreted as abstract position and momentum
determinations, hence the commutation relations that we impose on
them may be regarded as {\em abstract sheaf cohomological
Heisenberg uncertainty relations}. Of course, the formal analogy
with the 2nd canonical field-quantization of gravity and gauge
field theories stops here since, in glaring contrast to those two
scenaria, our scheme is manifestly background spacetime
manifold-free and it thus regards gravity as a pure ({\it ie},
external spacetime continuum unconstrained) quantum gauge theory.
In this way, gauge theory of the 3rd kind and 3rd-quantization
appear to go hand in hand.

We also emphasize the manifest {\em functoriality} of
3rd-quantization, and then we draw preliminary, albeit suggestive,
links between it and Mallios' ADG-based $K$-theoretic treatment of
geometric (pre)quantization and second quantization in
\cite{mall6,mall4}. Based on these $K$-theoretic smatterings, we
highlight close affinities between 3rd-quantized VEG and FYM, and
our ADG-based {\em finitary, causal and quantal VEG and FYM} in
\cite{malrap1,malrap2,malrap3,malrap4,rap5,rap7,rap11}. In the
concluding section (4), we summarize our findings and discuss
briefly the potential impact that 3rd-quantization may have on
certain outstanding (and persistently resisting resolution!)
problems in current and future QG research.
\section{Geometrically (Pre)quantized and Second Quantized Vacuum Einstein
Gravity and Free Yang-Mills Theories}

From the ADG-theoretic perspective, vacuum Einstein gravity (VEG)
is regarded and treated as a pure gauge theory, like the free
Yang-Mills theories (FYM)
\cite{mall1,mall2,mall3,malrap1,malrap2,malrap3,malrap4,rap5,rap7,rap11}.This
means that the sole dynamical variable in the theory is an
algebraic $\struc$-connection $\conn$ on a vector sheaf $\modl$,
the (Ricci scalar) curvature of which, $\ricci(\conn)$, obeys the
equation

\begin{equation}\label{eq1}
\ricci(\modl)=0
\end{equation}

\noindent The corresponding formalism has been coined `{\em
half-order formalism}' and it should be contrasted against
Einstein's original 2nd-order one, where the only gravitational
dynamical variable is the smooth spacetime metric $g_{\mu\nu}$.
Perhaps more importantly {\it vis-\`a-vis} current QG trends,,
ADG-gravity should be contrasted against the more recent
Ashtekar-Palatini 1st-order formalism \cite{ash}, in which
although the smooth connection assumes a more assertive and
physically significant role, thus pronouncing  more the
gauge-theoretic character of gravity, the metric is still present
in the guise of the smooth tetrad field $e_{\mu}$.

In ADG-gravity, the equation (\ref{eq1}) is obtained from varying
with respect to $\conn$ an Einstein-Hilbert action functional
$\eh$  on the affine space $\sconn_{\struc}(\modl)$ of
$\struc$-connections, which may be formally identified with the
configuration space in the theory \cite{malrap3}. This is in
contrast to both the 2nd and the 1st-order formalism in which
variation of the Einstein-Hilbert functional with respect to the
metric and, what amounts to the same, with respect to the {\it
vierbein} field respectively, produces (\ref{eq1}). Moreover, in
the 1st-order formalism, variation with the connection field
produces the auxiliary metric compatibility condition for the
connection

\begin{equation}\label{eq2}
\conn g=0
\end{equation}

\noindent By contrast, in ADG-gravity the
$\struc$-metric\footnote{Symbolized by $\rho$ in the theory
\cite{mall1,mall2,mall3,malrap3}.} is a physically secondary ({\it
ie}, a dynamically not primary) structure, while its compatibility
with the $\struc$-connection $\conn$ is optional to the theorist,
and certainly {\em not} deducible variationally from the dynamical
action, which is a functional of the connection
exclusively.\footnote{To be sure, in ADG-gravity the usual `{\em
metric compatibility of the connection}' condition (\ref{eq2})
above is still observed, but in the other way round.  That is to
say, {\em if} the theorist chooses to impose an $\struc$-metric
structure $\rho$ in the theory (:on $\modl$), then she might like
to make sure that this metric is compatible with ({\it ie}, it
respects) the $\struc$-connection $\conn$, which is the primary
dynamical structure on $\modl$. Thus, in ADG-gravity one talks
about the `{\em connection compatibility of the metric}', which is
equivalent to the following `{\em horizontality condition}' for
the connection on the tensor product vector sheaf
$\Hom_{\struc}(\modl
,\modl^{*})=(\modl\otimes_{\struc}\modl)^{*}=\modl^{*}\otimes_{\struc}\modl^{*}$
induced by the $\struc$-connection $\conn$ on $\modl$:
$\conn_{\Hom_{\struc}(\modl ,\modl^{*})}(\tilde{\rho})=0$, where
$\tilde{\rho}$ effectuates the canonical $\struc$-isomorphism
$\modl\stackrel{\tilde{\rho}}{\cong}\modl^{*}$
 between $\modl$ and its dual $\modl^{*}\equiv\Hom(\modl ,\struc):\simeq \Omg$ \cite{malrap3}. }

With \cite{mall1,mall2,malrap3,mall4} as reference guides to the
technical symbols, terms, their definitions and associated
construction details, direct comparison between (\ref{eq1}) and
the FYM equation

\begin{equation}\label{eq3}
\delta\stre(\conn)=0~~\mathrm{or}~~\Delta\stre(\conn)=0\footnote{Where
$\stre(\conn)$ is the curvature (:field strength) of the
Yang-Mills connection $\conn$, while $\delta$ and $\Delta$ the
ADG-versions of the usual coderivative and Laplacian differential
operators, respectively.}
\end{equation}

\noindent as well as between the Einstein-Hilbert action
functional $\eh$ and the Y-M one $\ym$ on the corresponding affine
spaces $\sconn_{\struc}(\modl)$ of $\struc$-connections on the
respective $\modl$s involved, shows what was mentioned in the
beginning, namely, that

\begin{quotation}
\noindent {\em from the ADG-theoretic vantage, VEG is a `pure'
gauge theory, like the FYM theory.}
\end{quotation}

\noindent A more glaring contrast between ADG-gravity and the
usual 1st and 2nd-order formulation of GR (both of which rely
mathematically on the CDG of $\smooth$-smooth pseudo-Riemannian
manifolds) is that it does not employ at all any background
geometrical locally Euclidean space (:differential manifold) to
formulate the theory differential geometrically. Rather, it relies
solely on purely algebraico-categorical (:sheaf-theoretic) means
to formulate its concepts and develop its constructions. It
follows that

\begin{quotation}
\noindent {\em the theory's physical semantics does not involve
any background spacetime continuum interpretation and its
associated `geometrical picturization'  either}.
\end{quotation}

\noindent  All that is involved in ADG-gravity---its fundamental,
`{\it ur}' element so to speak---is the ADG-gravitational field
$\field$, which is defined as a pair

\begin{equation}\label{eq4}
\field :=(\modl ,\conn)
\end{equation}

\noindent consisting of a {\em vector sheaf}---by definition, a
locally free $\struc$-module of finite rank $n$ on an in principle
arbitrary topological space $X$,\footnote{$\struc$ is a sheaf of
unital,  and associative {\em differential} (and not necessarily
functional, strictly speaking) $\mathbf{K}$-algebras
($\mathbf{K}=\mathbb{R}_{X},\com_{X}$: the constant sheaf of real
or complex numbers over $X$) called the {\em structure sheaf of
generalized arithmetics} (the terms `coefficients' and
`coordinates' are synonyms to `arithmetics'). By definition,
$\modl$ is locally a finite power of $\struc$:
$\Gamma(U,\modl):=\modl(U)\equiv\modl
|_{U}\simeq\struc(U)^{n}=\struc^{n}(U)$, $U$ open in $X$. At the
same time, $X$ is usually taken to be .} and a linear, Leibnizian
{\em sheaf morphism} $\conn$, the $\struc$-connection, acting on
$\modl$'s local sections in $\modl(U)$. This is a particular
instance of the general ADG-theoretic notion of a field $\field$
as a pair $(\modl ,\conn)$, which has been abstracted from the
usual conception of a field as a connection on a smooth principal
fiber (or its associated/representation vector)
bundle.\footnote{For example, the classical electromagnetic field
of Maxwellian electrodynamics is regarded as the pair
$({\mathcal{L}},\conn)$ consisting of a connection $\conn$ on a
line bundle ${\mathcal{L}}$ (:the associated bundle of the $U(1)$
principal fiber bundle of electrodynamics) \cite{manin}.
Analogously, the Maxwell field in ADG is defined as a connection
on a line {\em sheaf} $\lsh$ (:a vector sheaf of rank $1$)
\cite{mall1,mall2,mall4,malrap2,malrap3}.}

Due to the manifest absence in ADG-VEG of a smooth background
spacetime manifold,

\begin{quotation}

\noindent {\em the ADG-gravitational field can be regarded as an
external smooth spacetime unconstrained gauge system.}

\end{quotation}

\noindent This is another result supporting our claim that
ADG-gravity is a pure gauge theory. Moreover, this fact has
profound consequences for plausible quantization scenaria within
the ADG-framework as we shall argue in the sequel. For one thing,
while the usual notions of `space' and `time' are {\em not}
primary in ADG-field theory, they may still be thought of as being
`inherent' in the {\it ur}-concept of ADG-field. For example,
`space' may be thought of as being already effectively encoded in
$\struc$ ({\it eg}, Gel'fand duality)
\cite{mall6,malrap3,mall9,mall11,mall10,malrap4},\footnote{Much
like in the usual theory (:CDG), a differential manifold $M$ can
be derived from the structure sheaf $\struc\equiv\smooth_{M}$ of
germs of smooth ($\R$-valued) functions on it as the latter's
(real) Gel'fand spectrum.} while an abstract notion of `time'
(:`dynamical change' or `progression') is arguably already
inherent in the dynamical evolution equation (\ref{eq1}) for
$\modl$'s states (:local sections) on which the ADG-gravitational
$\struc$-connection field $\conn$ acts, via its (Ricci) curvature,
to change them dynamically.

In fact, as it must have already been transparent from the
exposition so far, one can adopt and adapt the entire gauge
field-theoretic conceptual jargon and technical machinery to
ADG-field theory, briefly as follows: one can cover the base
topological space $X$ by a system $\gauge$ of {\em local open
gauges} $U$ and relative to it consider {\em local gauge frames}
$e^{U}$ ($U\in X$ open) constituting local bases of
$\modl(U)$.\footnote{That is, any local section $s\in\modl(U)$ can
be expanded as a unique linear combination of the $n$ linearly
independent local sections in $e^{U}$, with coefficients in
$\struc$.} Then, in view of the aforementioned local isomorphism
$\modl(U)\simeq\struc^{n}(U) $, one identifies the natural gauge
(:structure) group sheaf (:principal sheaf
\cite{mall1,vas1,vas2,vas3mall4}) of the ADG-gauge field pair
$(\modl ,\conn)$ with

\begin{equation}\label{eq5}
\aut\modl(U)=\modl
nd\modl(U)^{\bull}=M_{n}(\struc(U))^{\bull}={\mathcal{G}}{\mathcal{L}}(n,\struc)(U)
\end{equation}

\noindent {\em the group sheaf of local automorphisms of $\modl$}.
This latter group sheaf effectuates in ADG-gravity the abstract
version of the Principle of General Covariance (PGC), since it is
the ADG-analogue of the Lie group $GL(4,\R)$ of general coordinate
transformations in the $4$-dimensional spacetime manifold based GR
\cite{malrap3,malrap4,rap5,rap7,rap11}. Moreover, the principal
sheaf $\aut\modl$ is the sheaf-theoretic ADG-analogue of the
`structure' group $\mathrm{Diff}(M)$ of the base differential
spacetime manifold of GR,\footnote{Indeed, by assuming
$\smooth_{X}$ as structure sheaf $\struc$ in the theory, $X$ can
be identified with a smooth manifold $M$ by Gel'fand duality as
briefly noted earlier, and then plainly, by definition:
$\mathrm{Aut}M\equiv\mathrm{Diff}(M)$.} it too effectuating a
`global' version of the PGC of GR in ADG-gravity. In turn, as
briefly mentioned before, $\modl$ may be regarded as the
associated ({\it alias}, representation) sheaf of the principal
sheaf $\aut\modl$, carrying a representation of the (local)
structure group $\mathrm{GL}(n,\struc)$ in its fibers ({\it
alias}, stalks). We may summarize graphically the above in the
following diagram, which we borrow from \cite{mall4}:\footnote{I
wish to thank Mrs Popi Mpolioti, of the Algebra and Geometry
Section, Maths Department, University of Athens, for giving me the
LaTeX graphics for this `categorical/commutative' diagram, and of
course to Tasos Mallios for permitting me to borrow and slightly
modify it from his latest book \cite{mall4}.}

{\small
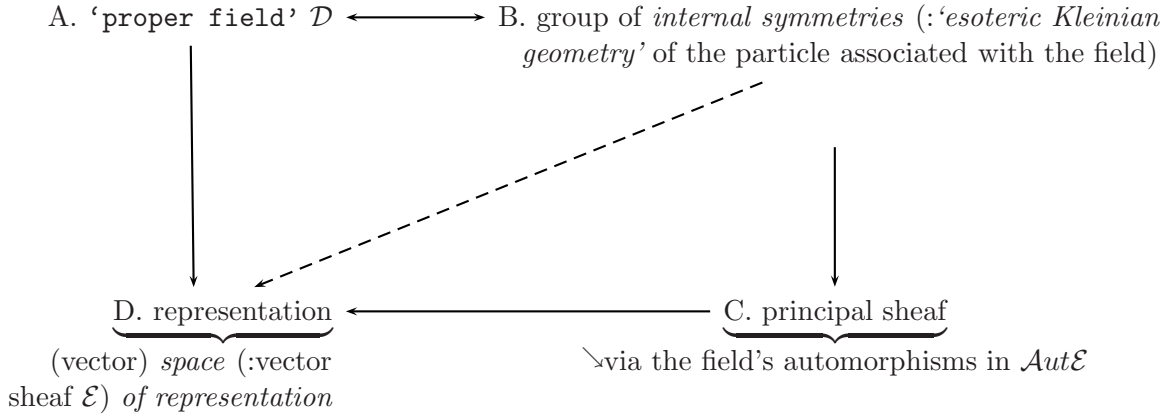
\[
\setlength{\arraycolsep}{1.1cm}%
\begin{array}{rcl}%
 \Rnode{a}{\text{A. \texttt{`proper field'}~$\conn$}}%
 &\Rnode{b}{\text{B. group of \emph{internal~symmetries}~\text{(:}\textit{`esoteric Kleinian }}}\\%
 \Rnode{c}{}
  & \Rnode{d}{\ \textit{geometry'}~\text{of~the~particle~associated~with~the~field)}}\\
 \Rnode{e}{}
  & \Rnode{f}{}\\
 \Rnode{g}{}
 & \Rnode{h}{}\\[2cm]%
 \Rnode{i}{\text{\qquad $\underbrace{\mathrm{D.~representation}}$}}
 & \Rnode{j}{\underbrace{\text{C. principal sheaf}}} \\
 \Rnode{k}{\text{(vector) \emph{space} (:\text{vector}}}
  & \Rnode{l}{\overset{\searrow}{} \text{via the field's automorphisms in $\aut\modl$}}\\
 \Rnode{m}{\text{sheaf}~\modl)~\emph{of representation}}
 & \Rnode{n}{}
\end{array}
\everypsbox{\scriptstyle}%
\psset{nodesep=5pt,arrows=<->}%
\ncline{a}{b}\taput{}%
\psset{nodesep=5pt,arrows=->}%
\ncline{a}{i}\tlput{}%
\ncline{h}{j}\tlput{}
\ncline[linestyle=dashed]{d}{i}%
\psset{nodesep=5pt,arrows=<-}%
\ncline{i}{j}\tbput{}%
\]
}

\noindent Note in the diagram above that by `\texttt{proper
field}' we refer to the connection part $\conn$ of the ADG-field
pair $\field =(\modl ,\conn)$. This separation and distinction
between the `proper field' ($\conn$) and the `total field'
$\field=(\modl ,\conn)$ may seem superfluous at first sight, but
it is of great semantic significance: namely, in ADG the field
proper is the connection $\conn$ existing `out there'
independently of us,\footnote{That is, independently of our
generalized measurements (:`coordinatizations') in $\struc$ and
associated geometrical representations by
$\modl\stackrel{\mathrm{loc.}}{=}\struc^{n}$.} which is then
geometrically materialized (:represented) by $\modl$ when we
coordinatize it by introducing a structure sheaf $\struc$ of our
choice. The deeper significance of this distinction will become
clearer in the next section where we introduce our third ADG-field
quantization scenario and we discuss the $\struc$-functoriality of
the ADG-gravitational dynamics (\ref{eq1}), of 3rd-quantization in
general, as well as the principle of ADG-field realism that this
functoriality entails.

It is fitting to stress at this point that, in the past
\cite{malrap3,malrap4,rap5,rap7,rap11} the purely gauge
field-theoretic ADG-perspective on gravity has been coined `{\em
gauge theory of the third kind}', due to the following features:

\begin{enumerate}

\item As noted earlier, the sole dynamical variable is the
algebraic $\struc$-connection $\conn$ (:$\frac{1}{2}$-order
formalism);

\item The scheme is manifestly local (:sheaf-theoretic) like the
current gauge theories of the second kind, and in
contradistinction to Weyl's original gauge theory of the first
kind, which was a global gauge (:scale) theory;

\item On the other hand, our ADG-gauge field theory is {\em not}
local in the usual sense of the modern-day gauge theories of the
second kind---{\it ie}, the gauge transformations (and symmetries)
are {\em not} localized  over an external, base (:background)
spacetime continuum (:manifold), since the latter does not exist
in our theory. All there is in our scenario is the dynamically
autonomous ADG-gravitational field $(\modl ,\conn)$, which does
not depend on a background spacetime manifold, and solely in terms
of which (and its curvature) the VEG dynamics is expressed as in
(\ref{eq1});

\item It follows that in our scheme, unlike the physical semantics
of nowadays gauge theories of the second kind, there is no
distinction between external (`spacetime') and internal (`gauge')
transformations (and dynamical symmetries). All our
transformations (and dynamical symmetries) are `internal'
(:gauge), taking place within the ADG-gravitational field $(\modl
,\conn)$, and are represented by $\aut\modl$.

\item The features above reveal an unprecedented {\em fundamental
dynamical autonomy} in ADG-gravity, which is part and parcel of
the theory's genuine {\em background spacetime manifold
independence}. Namely, all that `exists' and is of physical
significance in ADG-VEG (and FYM theories) is the autonomous
dynamical field $(\modl ,\conn)$, the law that it obeys/defines as
a differential equation proper (\ref{eq1})-(\ref{eq3}), and the
latter's `dynamical self-invariances' (:`autosymmetries') in
$\aut\modl$, without any reference to or dependence on an
extraneous (:externally imposed) structure ({\it eg}, background
spacetime manifold) to support that
autodynamics.\footnote{Elsewhere \cite{malrap3,malrap4,rap5}, we
have coined this Leibnizian monad-type of dynamical autonomy of
the ADG-field `dynamical holism', or even, `unitarity'. Like
Leibniz's monads, the ADG-fields possess their own `dynamical
entelechy', but at the same time unlike them, they are not
`windowless', as they can (dynamically) interact with each other.
Here, however, we have presented only the free field theories.}

\end{enumerate}

All the ideas above synergistically come to fruition when ADG is
applied for the geometric (pre)quantization and second
quantization of gauge theories, including gravity
\cite{mall1,mall2,mall5,mall6,malrap2,malrap3,mall4}. Indeed,
there $\modl$ is regarded as the Hilbert (or Fock) $\struc$-module
sheaf associated to the (principal) Klein group sheaf $\aut\modl$
of field automorphisms. The basic result there is the following
identification:

\begin{equation}\label{eq6}
\mathrm{local~quantum~particle~states~of~the~
field}\longleftrightarrow\mathrm{local~sections~of~\modl}\footnote{Notice
here the `{\em self-duality}' of the {\em total} ADG-field
$\field=(\modl ,\conn)$ in (\ref{eq5}): $\field$ has a `{\em
particle}' (:$\modl$) and a `\texttt{proper field}' (:$\conn$)
aspect, and it has thus been referred to as a `{\em particle-field
pair}' \cite{mall6,malrap3,mall4,malrap4}. In turn, this
self-duality of the total ADG-field pair $\field$ is an abstract
version of the usual `{\em quantum-theoretic duality}' between the
`{\em wave/momentum}' (here, $\conn$) and `{\em
particle/position}' (here, $\modl$) aspects of quanta, as we shall
argue in the next section in connection with our 3rd-quantization
of ADG-fields. There, we shall see that ADG-fields are `{\em
quantum self-dual}' or `{\em self-complementary}' entities (in the
quantum sense of `complementarity' due to Bohr)---another feature
pointing to their (quantum dynamical) autonomy alluded to
earlier.}
\end{equation}

\noindent which is then carried further to conclude that

\begin{quotation}
\noindent {\em every elementary field is geometrically
(pre)quantizable and second quantizable, that is to say, it admits
a prequantizing vector (Hilbert $\modl\equiv\hil$), or a second
quantizing vector (Hilbert-Fock
$\sum_{n\in\mathbb{Z}_{+}}\otimes_{\struc}^{n}\modl$), sheaf as
the representation state space of its identical particle-quanta}.
In particular, the {\em spin-statistics connection} of the usual
spacetime manifold and CDG-based QFTs of matter is observed in
that {\em local quantum particle states of boson (:integer spin)
fields are represented by vector sheaves of rank $n=1$ (:line
sheaves), while those of fermion (:half-integer spin) fields by
sections of vector sheaves of rank $n>1$}.
\end{quotation}

In view of these results, we can now claim that

\begin{quotation}
\noindent (\ref{eq1}) {\em is a geometrically (pre)quantized and
second quantized version of the VEG equations, for a suitable
choice\footnote{Effectively, for an $\struc$ chosen by the
theorist, since $\modl$ is locally a finite power of $\struc$.} of
representation sheaf $\modl$ for the free gravitational quanta}.
The same holds for (\ref{eq3}) and the free (`bare') gauge bosons
carrying the other three fundamental gauge forces. On the other
hand, matter quanta ({\it eg}, electrons) have connections ({\it
eg}, the Dirac operator) defined on vector sheaves of rank greater
than $1$ (:Grassmannian $\struc$-module sheaves).
\end{quotation}

In closing this section, we note as an {\it addendum} that when
for example $\struc$ is taken to be the  $C^{*}$-algebra sheaf of
germs of continuous $\com$-valued functions on a compact Hausdorff
topological space $X$ ({\it eg} a compact $C^{0}$-manifold), the
Kleinian endomorphism algebra sheaf $M_{n}(\struc)(U)$ of the
field may be regarded as the field's `noncommutative geometry'
{\it \`a la} Connes \cite{connes,block}. The commutative
coordinate functions in $\struc$ are promoted to `noncommutative'
ones, which now represent the field's intrinsic (dynamical)
self-transmutations (:endo/automorphisms) in
$M_{n}(\struc)(U)\equiv\modl nd\modl |_{U}$. This observation will
prove to be very important in the next section where we insist
that the Heisenberg-type of canonical commutation relations
defining third ADG-field quantization should close within $\modl
nd\modl$---the noncommutative (dynamical) Kleinian `auto-geometry'
of the ADG-field `in-itself'. The latter we may metaphorically
call `{\em quantum field foam}' as it is the structural quality of
the ADG-field that gives it its `foamy', `fuzzy', dynamically
ever-fluctuating character.

\section{Third Quantization of Gravity and Yang-Mills Theories}

One can carry the quantum physical interpretation of the ADG-gauge
field pair $(\modl ,\conn)$ even further and envisage a
canonical-type of field quantization scenario along sheaf
cohomological lines in the following physically intuitive and
mathematically heuristic way:\footnote{The arguments to be
presented below are conceptual, informal and tentative, and should
await a more formal and mathematically rigorous exposition. We
shall do this in a forthcoming paper \cite{rap13} (see the
declaration at the end).} we noted earlier that local sections of
$\modl$ represent local quantum-particle states of the ADG-VEG
field, while $\conn$ acts on them (via its curvature) to change
them dynamically according to (\ref{eq1}).

\begin{quotation}
\noindent We can thus, continuing our anticipatory remarks in
footnote 19, heuristically and intuitively interpret the local
sections of $\modl$ as abstract {\em position} or {\em coordinate}
determinations\footnote{After all, $\modl$ is locally of the form
$\struc^{n}$, and as noted earlier, $\struc$ represents our
abstract (local) coordinatizations (:local coordinate
determinations or `measurements') of the proper ADG-field $\conn$.
In turn, $\modl$ is the carrier ({\it alias}, representation)
space of the \texttt{proper field} $\conn$.} of the
(particle-quanta of the) field; while, as befits the (generalized)
differential operator $\conn$, interpret its effect/action on
those position states as an abstract {\em momentum}
map.\footnote{After all, momentum is normally perceived as a
(`rate' of) change of position. Moreover, it must be noted here
parenthetically that since the topological base space $X$ plays no
important role in the differential geometric mechanism of the
theory, but it merely serves as a scaffolding or `surrogate space'
for the sheaf-theoretic localization of the ADG-fields (for
instance, since the gravitational dynamics (\ref{eq1}) is
expressed categorically as an equation between $\struc$-sheaf
morphisms such as the curvature of the connection, which is an
$\otimes_{\struc}$-tensor, it `sees through' $X$; see remarks on
$\struc$-functoriality later in this section), there is {\em no}
notion of {\em tangent space} to it in ADG. It follows that the
local sections of $\modl$ should {\em not} be interpreted as {\em
tangent vectors} like in the usual theory (:CDG) of vector bundles
over a smooth base manifold ({\it eg}, the tangent bundle $TM$);
hence the theory does {\em not} accommodate {\em derivations},
which are normally defined as maps
$\mathrm{Der:}\,\struc\rightarrow\struc$ and are represented by
tangent vectors to the continuum. The abstract momentum maps noted
above are {\em not} derivations in the usual (:CDG, fiber
bundle-theoretic) sense.}
\end{quotation}

Now, since the ADG-fields $(\modl ,\conn)$ are dynamically
self-supporting, autonomous entities as we emphasized earlier;
moreover, since they are `self-dual' as it was anticipated in
footnote 19,

\begin{quotation}
\noindent {\em a possible quantization scenario for them should
involve solely their two constitutive parts, namely, $\modl$ and
$\conn$, without recourse to/dependence on extraneous structures
({\it eg}, a base spacetime manifold) for its (physical)
interpretation}.
\end{quotation}

Thus, in what formally looks like a canonical quantization-type of
scenario,

\begin{quotation}
\noindent {\em we envisage abstract non-trivial local commutation
relations between the abstract position} (:$\modl$) {\em and
momentum} (:$\conn$) {\em aspects of the ADG-fields}.
\end{quotation}

To this end, we recall that

\begin{quotation}
\noindent {\em there are certain local (:differential) forms that
uniquely characterize sheaf cohomologically the vector sheaf
$\modl$ and the connection $\conn$ parts of the ADG-fields}
$(\modl ,\conn)$
\end{quotation}

Thus, the basic intuitive idea here is to identify the relevant
forms and then posit non-trivial commutation relations between
them. Moreover, for the sake of the aforementioned `{\em dynamical
ADG-field autonomy}', we would like to require that

\begin{quotation}
\noindent the envisaged commutation relations should not only
involve just the two components ({\it ie}, $\modl$ and $\conn$) of
the total ADG-fields $\field$, but they should also somehow  {\em
close within the $\field$s themselves}---{\it ie}, the result of
their commutation relations should not take us `outside' the total
ADG-field structure (and its `auto-transmutations'), which anyway
is the only dynamical structure involved in our
theory.\footnote{This loosely reminds one of the theoretical
requirement for algebraic closure of the algebra of quantum
observables in canonical QG, with the important difference however
that the $\mathrm{Diff}(M)$ group of the external (to the
gravitational field) spacetime manifold must also be considered in
the constraints, something that makes the desired closure of the
observables' algebra quite a hard problem to overcome
\cite{thiem2}. Later on, we shall return to discuss certain
difficult problems that $\mathrm{Diff}(M)$ creates in various QG
approaches, as well as how its manifest absence in ADG-gravity can
help us bypass them totally. For, recall that from the
ADG-perspective gravity is an external (:background) spacetime
manifold unconstrained pure gauge theory (:of the 3rd kind).}
\end{quotation}

Keeping the theoretical requirements above in mind, we recall from
\cite{mall1,mall2,mall5,malrap2,mall4} two  important sheaf
cohomological results:

\begin{enumerate}

\item That, sheaf cohomologically, the vector sheaves $\modl$ are
completely characterized by a so-called {\em coordinate
$1$-cocycle} $\phi_{\alpha\beta}\in Z^{1}(\gauge
,{\mathcal{G}}{\mathcal{L}}(n,\struc))$ associated with any system
$\gauge$ of local gauges of $\modl$. Intuitively, this can be
interpreted in the following Kleinian way: since any (vector)
sheaf is completely determined by its (local) sections,\footnote{A
basic motto (:fact) in sheaf theory is that ``{\em a sheaf is its
sections}'' \cite{mall1}. If we know the local data (:sections),
we can produce the whole sheaf space by restricting and collating
them relative to an open cover $\gauge$ of the base topological
space $X$. This is the very process of `sheafification' (of a
preasheaf) \cite{mall1}.} one way of knowing the latter is to know
how they transform---in passing, for example, from one local gauge
($U_{\alpha}\in\gauge$) to another ($U_{\beta}\in\gauge$), with
$U_{\alpha}\cap U_{\beta}\neq\emptyset$ and $\gauge$ a chosen
system of local open gauges covering $X$.\footnote{In particular,
$\phi_{\alpha\beta}$ can be locally expressed as the $\struc
|_{U_{\alpha\beta}}$-isomorphism:
$\phi_{\alpha}\circ\phi_{\beta}^{-1}\in\aut_{\struc_{\alpha\beta}}
(\struc^{n}|_{U_{\alpha\beta}})=\mathrm{GL}(n,\struc(U_{\alpha\beta}))={\mathcal{G}}{\mathcal{L}}(n,\struc)(U_{\alpha\beta})$,
in which expression the familiar local coordinate transition
(:structure) functions appear. Hence, also the `natural' structure
(:gauge) group sheaf
$\aut\modl={\mathcal{G}}{\mathcal{L}}(n,\struc)$ of $\modl$
arises.}  To know something ({\it eg}, a `space') is to know how
it transforms, the fundamental idea underlying Klein's general
conception of `geometry'. The bottom-line here is that the
characteristic cohomology classes of vector sheaves $\modl$ are
completely determined by $\phi_{\alpha\beta}$; write:

\begin{equation}\label{eq7}
[\phi_{\alpha\beta}]\in
H^{1}(X,{\mathcal{G}}{\mathcal{L}}(n,\struc))=\lim_{\overrightarrow{\;\;\gauge\;\;}}H^{1}(\gauge,
{\mathcal{G}}{\mathcal{L}}(n,\struc))
\end{equation}

\noindent where the $\gauge$s, normally assumed to be {\em locally
finite open coverings of} $X$
\cite{mall1,mall2,malrap1,malrap2,malrap3}, constitute a {\em
cofinal subset} of the set of all proper open covers of
$X$.\footnote{An assumption that in the past has proven to be very
fruitful in applying ADG to the formulation of a locally finite,
causal and quantal VEG
\cite{malrap1,malrap2,malrap3,rap5,rap7,rap11}. We will use it in
our $K$-theoretic musings in the sequel, but provisionally we note
that the direct (:inductive) limit depicted in (\ref{eq7}) above
is secured by the `cofinality' of the set of finitary (:locally
finite) open coverings of $X$ that we choose to employ
\cite{sork0,rapzap1,rapzap2,rap1,rap2,malrap1,malrap2,malrap3,rap5,rap7}.}
{\it In toto}, we assume that $\phi_{\alpha\beta}$ encodes all the
(local) information we need to determine the local
quantum-particle states of the field in focus ({\it ie}, the local
sections of $\modl$).

\item On the other hand, it is well known that {\em locally
$\conn$ is uniquely determined by the so-called `gauge potential'
$\omega$}, which is normally ({\it ie}, in CDG) defined as a Lie
algebra (:vector) valued $1$-form. Correspondingly, in ADG
$\omega$ is seen to be an element of
$M_{n}(\Omega(U))=M_{n}(\Omega)(U)=\Omega(\modl
nd\modl)$,\footnote{Note that, as also mentioned earlier in
footnote 9, in ADG by definition one has: $\Omg
:=\modl^{*}:={\mathcal{H}}om_{\struc}(\modl  ,\struc)$. That is,
the $\struc$-module sheaf $\Omg$ of abstract differential
$1$-forms is dual to the vector sheaf $\modl$, much like in the
classical theory (:CDG of $\smooth$-manifolds) where differential
forms (:cotangent vectors) are dual to tangent vectors, although
again as noted earlier in footnote 23, in ADG the epithet
`(co)tangent' is meaningless due to the manifest absence of an
operative background space(time) of any kind (especially, of a
base manifold).} thus it is called the {\em local
$\struc$-connection matrix $(\omega_{ij})$ of $\conn$, with
entries local sections of $\modl^{*}=\omg$}. In turn, this means
that locally $\conn$ splits in the familiar way, as follows:

\begin{equation}\label{eq8}
\conn =\partial +\omega
\end{equation}

\noindent where $\partial$ is the usual `inertial' (:flat)
differential\footnote{In ADG, $\partial$, like $\conn$, is defined
as a linear, Leibnizian $\mathbf{K}$-sheaf morphism $\partial :\;
\struc\rightarrow\omg$, thus it is an instance of on
$\struc$-connection; albeit, a  {\em flat} one
(:$\curv(\partial)=0$)
\cite{mall1,mall2,malrap1,malrap2,malrap3}.} and $\omega$ the said
gauge potential. In ADG-gravity, the \texttt{proper field} $\conn$
as a whole (:`globally') represents the {\em gravito-inertial
field}, but locally it can be separated into its inertial
(:$\partial$) and gravitational (:$\omega$) parts.\footnote{For
more discussion on the physical meaning of this local separation
of the \texttt{proper field} $\conn$ into $\partial$ and $\omega$,
see footnote 34 below.}

\end{enumerate}

Thence, the envisaged sheaf cohomological canonical
quantization-type of scenario for the total ADG-fields
$\field=(\modl ,\conn)$ rests essentially on positing the
following non-trivial abstract Heisenberg-type local commutation
relations between (the characteristic forms that completely
characterize) $\modl$ (:abstract {\em position} states) and
$\conn$ (:abstract {\em momentum} operator). Thus, heuristically
we posit:

\begin{equation}\label{eq9}
[\phi_{\alpha\beta},\partial
+\omega_{ij}]_{U_{\alpha\beta}}=[\phi_{\alpha\beta},\partial
]_{U_{\alpha\beta}}+[\phi_{\alpha\beta},\omega_{ij}]_{U_{\alpha\beta}}
\end{equation}

\noindent stressing also that

\begin{quotation}
\noindent the local commutation relations in (\ref{eq9}) above are
well defined, since they effectively {\em close within the
noncommutative $(n\times n)$-matrix Klein-Heisenberg algebra}
$\modl
nd\modl(U_{\alpha\beta})=M_{n}(\struc(U_{\alpha\beta}))=M_{n}(\struc)(U_{\alpha\beta})$
of the field's endomorphisms---the field's `noncommutative
Kleinian geometry' we mentioned at the end of the last section
(:quantum field foam).
\end{quotation}

\noindent This `{\em algebraic closure}' is in accord with the
theoretical requirement we imposed  earlier, namely that,

\begin{quotation}
\noindent the abstract, Heisenberg-like, canonical quantum
commutation relations between the two components $\modl$ and
$\conn$ of the ADG-fields should not take us outside the fields,
but should rather {\em close} within them.\footnote{Here, one
could envisage an abstract Heisenberg-type of algebra freely
generated (locally) by $\phi$ (:abstract position) and $\omega$
(:abstract momentum), modulo the (local) commutation relations
(\ref{eq9}). Plainly, it is a subalgebra of $\modl nd\modl(U)$,
but deeper structural investigations on it must await a more
complete and formal treatment \cite{rap13}.}
\end{quotation}

\noindent Indeed, $\modl nd\modl$ {\em is precisely the algebra
sheaf of internal/intrinsic (dynamical) self-transmutations of the
(quantum particle states of the) field}---by definition, the
$\modl$-endomorphisms in ${\mathcal{H}}om_{\struc}(\modl ,\modl)$
(:quantum field foam). This is another aspect of the quantum
dynamical autonomy of ADG-fields:

\begin{quotation}
\noindent the $\modl$ (:abstract point-particle/position) part of
the ADG-field is `{\em complementary}', in the quantum sense of
`complementarity', to $\conn$ (:abstract field-wave/momentum).
Thus, {\em the total ADG-fields $\field$ are `quantum self-dual'
entities} \cite{malrap3,malrap4,rap5,rap11}, as we anticipated in
footnote 19.\footnote{From our abstract perspective, the de
Broglie-Schr\"odinger wave-particle duality is almost tautosemous
with the Bohr-Heisenberg momentum-position complementarity.}
\end{quotation}

Furthermore, by choosing $\phi_{ab}=\phi_{ab}^{in}$\footnote{The
superscript `$in$' stands for `{\em inertial}', and it represents
a choice (:{\em our} choice!; see next footnote) of a local
change-of-gauge
$\phi^{in}_{\alpha\beta}\in\gl(n,\struc)_{\alpha\beta}\equiv\Gamma(U_{\alpha\beta},\gl(n,\struc)$
that would take us to a locally inertial frame of $\modl$ over
$U_{\alpha\beta}\subset X$.} so that $\omega$ is `gauged
away'---{\it ie}, by setting $\omega=0$,\footnote{This is an
analogue of the Equivalence Principle (EP) of GR in ADG-gravity,
corresponding to the local passage to an `inertial frame' (:one
`covarying' with the gravitational field; {\it eg}, recall
Einstein's free falling elevator {\it gedanken} experiment) in
which the curved gravito-inertial $\conn$ in (\ref{eq8}) reduces
to its flat `inertial' $\struc$-connection part $\partial$
\cite{mall1,mall2,malrap1,malrap2,malrap3}. This just reflects the
well known fact that {\em GR is locally SR}, or conversely, that
when SR is localized ({\it ie}, `gauged' over the base spacetime
manifold) it produces GR (equivalently, the curved Lorentzian
spacetime manifold of GR is locally the flat Minkowski space of
SR). {\it In summa}, gravity (:$\omega$) has been locally gauged
away, and what we are left with is the inertial action $\partial$
of the ADG-gravitational field $\conn$. It must be also stressed
here that the choice of a locally inertial frame, like all gauge
choices, is an externally imposed constraint in the
theory---`externally', meaning that it is {\em we}, the external
(to the field) experimenters/theoreticians (`observers') that we
impose such constraints on the field ({\it ie}, we make choices
about what aspects of the field we would like to single out and,
ultimately, observe/study).} reduces (\ref{eq9}) to (omitting the
local open gauge indices/subscripts `$\alpha ,\beta$'):

\begin{equation}\label{eq10}
[\phi^{in} ,\partial ]=\phi^{in}\circ\partial
-\partial\circ\phi^{in}
\end{equation}

\noindent Moreover, since we are sheaf cohomologically guaranteed
that $\partial\circ\phi=0$ globally, which is tantamount to the
very {\em existence} of an $\struc$-connection $\conn$ (globally)
on $\modl$ \cite{mall1,mall2,mall4},\footnote{This essentially
corresponds to the fact that the coordinate $1$-cocycle
$\phi_{\alpha\beta}\in Z^{1}(\gauge ,\Omega)$ is actually a {\em
coboundary} (:a closed form), belonging to the zero cohomology
class $[\partial\phi_{\alpha\beta}]=0\in H^{1}(X,M_{n}(\Omega))$,
which in turn guarantees the existence of an $\struc$-connection
on $\modl$ as the so-called {\em Atiyah class} $\mathfrak{a}$ of
$\modl$ vanishes
(:$\mathfrak{a}(\modl):=[\partial\phi_{\alpha\beta}]=0$)
\cite{mall1,mall2,mall4}.} (\ref{eq10}) further reduces to:

\begin{equation}\label{eq11}
[\phi^{in} ,\partial ]=\phi^{in}\circ\partial
\end{equation}

\noindent Now, a heuristic physical interpretation can be given to
(\ref{eq11}) if we consider its effect (:action) on a local
section
$s\in\modl_{\alpha\beta}:=\modl(U_{\alpha\beta})\equiv\modl
|_{U_{\alpha\beta}}$:

\begin{equation}\label{eq12}
[\phi^{in} ,\partial
](s)=(\phi^{in}\circ\partial)(s)=\phi^{in}(\partial s)
\end{equation}

\noindent (\ref{eq12}) designates the inertial dynamical action of
$\conn$ ({\it ie}, the action of its locally flat, inertial part
$\partial$) on (an arbitrary) $s$, followed by the gauge
transformation of $\partial s$ to an inertial frame
$e_{in}^{U_{\alpha\beta}}\subset\modl_{\alpha\beta}$ `{\em
covarying}' with the inertio-gravitational field. It expresses
what happens to a `vacuum graviton state' $s$ when it is first
acted upon\footnote{Recall that we are considering only {\em
vacuum} gravity, in which the non-linear gravitational field
`couples' solely to itself(!)} by the inertial part of the
\texttt{proper} ADG-gravitational field $\conn$ and
then\footnote{The sequential language used here should not be
interpreted in an temporal-operational sense---as it were, as
`operations carried out sequentially in time'.} to an inertial
frame that in a sense `{\em covaries}' with the said inertial
change $\partial$ of $s$.

Perhaps one can get a more adventurous (meta)physical insight into
(\ref{eq12}) by defining the {\em uncertainty operator} $\unc$ as

\begin{equation}\label{eq13}
\unc :=\phi^{in}\circ\partial\in\modl nd\modl
\end{equation}

\noindent and by delimiting all the quantum-particle (:abstract
position) states of the field (:local sections of $\modl$) that
are annihilated by it. Intuitively, these are formally the local
`{\em classical-inertial}' states

\begin{equation}\label{eq14}
\modl^{cl}_{U}:=\mathrm{span}_{\mathbf{K}}\{ s\in\modl(U) :\;
\unc(s)=0\}=:\mathrm{ker}(\unc)
\end{equation}

\noindent for which the abstract sheaf cohomological Heisenberg
uncertainty relations (\ref{eq10}) vanish. Plainly,
$\modl^{cl}(U)$ is a $\mathbf{K}$-linear subspace of
$\modl(U)$---the kernel of $\unc$.

On top of the above, intuitively it makes sense to assume that
$\unc$ is a `projector'---a primitive idempotent (:projection
operator) locally in $\modl nd\modl$ ({\it ie}, in
$M_{n}(\struc(U))$)---since the `gedanken' operation of `{\em
inertially covarying with a chosen local inertial frame}' must
arguably be idempotent.\footnote{After all, `{\em inertially
covarying the inertial state leaves it inertially covariant}'. Or,
to use a famous Einstein `{\it gedanken} metaphor': `{\em jumping
on a light-ray (in order to ride it) twice, simply leaves you
riding it}'(!)} This means that $\unc^{2}=\unc$, so that $\unc$
separates (chooses or projects out) the `classical'
($\mathrm{eigen}_{0}(\unc)\equiv\mathrm{ker}(\unc)$) from the
quantum ($\mathrm{eigen}_{1}(\unc)$) local quantum
gravito-inertial states. A formal reason why we chose $\unc$ to be
a projection operator will become transparent in our $K$-theoretic
musings a little bit later.

Finally, we would like to ask {\it en passant} here the following
highly speculative question:

\begin{quotation}
\noindent Could the generation/emergence of
(inertio-gravitational) mass be somehow accounted for by a
(spontaneous)  symmetry breaking-type of mechanism, whereby, the
dynamical automorphism group $\aut\modl$ of the ADG-gravitational
field $(\modl ,\conn)$ reduces to its subroup that leaves
$\mathrm{ker}(\unc)$ invariant? Alternatively intuited, could the
emergence of inertio-gravitational mass be thought of as the
result of some kind of `quantum anomaly' of 3rd-quantized
VEG?\footnote{The epithet `quantum' adjoined to `anomaly' is
intended to distinguish the effect intuited above from the usual
anomalies. A `quantum anomaly' is the `converse' of an anomaly in
the usual sense, in that what was a symmetry of the {\em quantum}
theory (:an element of $\aut\modl$ in our case) ceases to be a
symmetry of the `classical domain' of our theory
(:$\mathrm{ker}(\unc)$). Let it be stressed that the emergence of
gravito-inertial `mass' in the sense intuited here has a truly
relational (:algebraic) and `global' flavour reminiscent of Mach's
ideas: `global' gravitational field symmetries in $\aut\modl$ are
locally reduced to inertial ones, and sheaf theory's ability to
interplay between local and global comes in handy in this respect
\cite{malrap4}. (See further remarks below on how sheaf theory
allows us to go from `local' to `global', and vice versa.)}
\end{quotation}

\subsection*{Functoriality: the quintessence of 3rd-quantization}

We commence this subsection by noting that in ADG-gravity the
notion of {\em functoriality} plays a very significant role and
has a very precise physical interpretation in the theory, with
significant, we believe, implications for certain current and
future QG research issues:

\begin{quotation}

\noindent In our scheme, functoriality pertains to {\em
$\struc$-functoriality} of the VEG and the FYM dynamics
(\ref{eq1})-(\ref{eq3}). This means that the geometrically
(pre)quantized, 2nd-quantized and, ultimately, 3rd-quantized VEG
and FYM dynamical equations are not `perturbed' at all by our acts
of coordinatization (:`measurements') encoded in the $\struc$ that
{\em we} choose up-front to employ as structure sheaf of
generalized arithmetics (:coordinates).\footnote{This choice of
ours may be understood as follows:  it is {\em we} that choose an
$\struc$ to coordinatize the dynamical connection \texttt{field
proper} $\conn$ and then represent it as acting on the vector
sheaf $\modl$. The latter, which is locally a finite power of
$\struc$, is the representation (:associated) sheaf of the group
sheaf $\aut\modl$ of dynamical self-transmutations
(:automorphisms) of the field, and it can thus be regarded as the
`{\em carrier space}' (for the action) of $\conn$.} Plainly, this
is so, because both the VEG curvature and the FYM field strength
involved in (\ref{eq1}) and (\ref{eq3}) respectively are
$\struc$-morphisms (:{\it alias},
$\otimes_{\struc}$-tensors).\footnote{Where $\otimes_{\struc}$ is
the {\em homological tensor product functor} between the relevant
categories involved (mainly, the category of $\struc$-modules).
$\otimes_{\struc}$-tensors  are the `geometrical objects' in our
theory as, in a Kleinian sense, they are left invariant under
$\aut_{\struc}\modl$.}
\end{quotation}

\noindent Concerning ADG-VEG in particular, $\struc$-functoriality
is an abstract version of the PGC of  the manifold based GR
\cite{malrap3,rap5,rap7,rap11,malrap4}.

At this point, before we go into discussing functoriality {\it
vis-\`a-vis} prequantization, 1st, 2nd and 3rd-quantization, we
would like to dwell for a while on how $\struc$-functoriality in
ADG-VEG may evade two apparently insurmountable (by CDG-means)
problems in classical and quantum GR.

The PGC of GR, when mathematically modelled after the
$\mathrm{Diff}(M)$ group of the $\smooth$-smooth spacetime
manifold based (and, {\it in extenso}, CDG-founded) GR, creates
serious problems in both the classical and the quantum theory,
briefly as follows:

\begin{enumerate}

\item Traditionally, gravitational singularities are supposed to
be a problem of the classical field theory of gravity (:GR).
There, the PGC appears to  come into conflict with the very
existence of singularities to the extent that until today there is
no unanimous agreement on (or anyway, a clear-cut definition of)
what is a singularity in the theory \cite{geroch,clarke3,clarke4}.
Granted that singularities are built into the differential
manifold that {\em we} assume up-front to model `spacetime' in
GR,\footnote{That is to say, singularities are {\it loci} in the
spacetime continuum $M$ in the vicinity of which some smooth
function component of the smooth metric
$\otimes_{\struc\equiv\smooth_{M}}$-tensor solution of the
Einstein equations (:$g_{\mu\nu}$) appears to blow up
uncontrollably without bound, and the associated differential
equations of Einstein appear to `break down' in one way or another
\cite{clarke3,clarke4,rap5,malrap4}. Thus, from the ADG-theoretic
perspective, smooth gravitational  singularities are inherent in
the coordinate structure sheaf  $\struc\equiv\smooth_{X}$ that we
assume up-front in GR, which is tantamount to our {\it a priori}
assumption of a base differential manifold $M$ in that CDG-based
classical field theory of gravity (Gel'fand duality).} it is
hardly surprising, especially in view of S1, that the usual
manifold based Analysis comes short of resolving or evading them
completely \cite{clarke3,clarke4,rap5,malrap4}. For, to tackle
singularities, we are in the first place using a Differential
Calculus (:CDG) that is vitally dependent on a background smooth
manifold that is carrying the very singularities we are trying to
resolve. There appears to be no way out of this vicious circle as
long as we insist on doing GR by manifold based CDG-means. To
paraphrase and extend  a well known quote of Peter Bergmann :

\begin{quotation}
\noindent it is not surprising that GR ``{\em carries the seeds of
its own destruction}'' \cite{berg} in the form of singularities,
since, the manifold based CDG employed to develop the classical
field theory of gravity already carries in its foundation, its
soil as it were (:$M\leftrightarrow\smooth(M)$)\footnote{Recall
Gel'fand duality: a differential manifold $M$ {\em is} the
topological algebra $\smooth(M)$ of smooth functions on it
\cite{mall0}.} those very singular germs (pun intended).
\end{quotation}

\noindent {\it Mutatis mutandis} then for the structure group of
the underlying spacetime manifold: it is not surprising that
$\mathrm{Diff}(M)$ clashes with our attempts at giving a clear-cut
`definition' of gravitational singularities. Plainly, at the basis
of the aforesaid vicious self-referential problem lies $M$, so
that what behooves us is to ask whether there is an alternative
differential calculus---one that does not depend at all on a base
manifold, yet one that can reproduce all the constructions and
results of the manifold based CDG, if we wished to---by which we
can view (and actually do!) gravity as a field theory.

Of course, for us this is a rhetorical question since we have ADG.
By ADG-means we have completely evaded singularities of all sorts
\cite{malros1,malros2,malros3,mall3,rap5,malrap4}, on the one hand
by doing away with a base differential spacetime manifold, while
on the other, by `absorbing' singularities in the structure sheaf
$\struc$ of generalized arithmetics and by formulating the VEG
differential equations {\em functorially} in terms of
$\struc$-sheaf morphisms. In this way singularities are {\em not}
seen to be sites where the Einstein equations break down
differential geometrically speaking, or where any sort of
unphysical infinity (in the usual analytical sense) is involved.
This must be attributed simply to the fact that ADG divests
Calculus from, to use another famous quote now of George Birkhoff
\cite{birkhoff}, ``{\em the glittering trappings of
Analysis}''---arguably, our being trapped into the aforementioned
vicious circle reflecting  our theoretical `imprisonment'  into
the background (spacetime) manifold that we assumed in the first
place(!) By breaking free from the background spacetime manifold
$M$, we totally bypass singularities, while the PGC of GR ceases
to be represented by $\mathrm{Diff}(M)$, but purely
algebraico-categorically, by $\aut_{\struc}\modl$ ---the
ADG-gravitational field's `autosymmetries'.

In addition to the above, we note that the $\struc$-functoriality
of the ADG VEG field dynamics, which corresponds to an abstract
version of the PGC of the manifold and CDG-based GR, can be
readily generalized further by categorical means to what has been
coined elsewhere the {\em Principle of Algebraic Relativity of
Differentiability} (PARD)
\cite{rap5,malrap4,rap7,rap11},\footnote{This is a generalization
of how the PGC of GR was seen to be a direct consequence of
Einstein's original Principle of Relativity maintaining that the
field law of gravity should hold in any coordinate system
\cite{einst1,einst3}.} as follows:

\begin{quotation}
\noindent Since, from the ADG-theoretic perspective, {\em all
differential geometry boils down to $\struc$}
\cite{mall1,mall2,mall4,malrap3,malrap4,rap5,rap7}---{\it ie}, all
differential geometry (indeed, the entire {\it aufbau} of ADG)
rests on the algebra (sheaf) of `{\em differentiable functions}'
that {\em we} assume/employ up-front (as coordinates) in the
theory\footnote{This `aphorism', {\it ie}, that all DG rests on
(our choice of) $\struc$, hinges on Mallios' fundamental
observation that the notion of differential $\partial$ ({\it viz.}
connection $\conn$)---arguably, {\em the} basic concept with which
one can actually talk about (and do!) {\em differential}
geometry---is vitally dependent on what (algebras of)  functions
we declare and assume up-front as being `differentiable'. These
functions then provide us with the differential operators we need
in order to do DG. Recall for example the very definitions of
$\partial$ and $\conn$ in ADG: $\partial :\,
\struc\rightarrow\Omg$ and $\conn :\, \modl\rightarrow\Omg$
\cite{mall1,mall2,malrap3}, both of which have effectively
$\struc$ as their domain of definition.}---while at the same time
the ADG-gravitational dynamics is $\struc$-functorial, any change
in ({\em our} choice of) structure sheaf
$\struc$\footnote{Essentially, any choice of what {\em we}
(`arbitrarily') perceive (and define!) as `{\em differentiable
functions}'. Such changes are entirely up to the theorist (and,
{\it in extenso}, to the observer or experimenter), if for
instance she wishes to consider another, more suitable to the
physical situation/problem she encounters, algebra of coordinate
functions in which to absorb a singularity that she confronts.}
should not affect ({\it ie}, at least it should leave
`form-invariant')  the ADG-VEG field dynamics (\ref{eq1}).
\end{quotation}

\noindent Categorically speaking, this gives rise to a {\em
natural transformation-type of functors} \cite{maclane1} between
the categories involved, which can be depicted short-handedly by
the following commutative diagram which we borrow almost intact
from \cite{malrap4,rap7}:

\begin{equation}
\bfig
\putsquare<1`1`1`1;900`900>(900,900)[\struc_{1}(:\modl_{1}\stackrel{\mathrm{loc.}}{\simeq}\struc_{1}^{n})`\ricci(\modl_{1})=0
`\struc_{2}(:\modl_{2}\stackrel{\mathrm{loc.}}{\simeq}\struc_{2}^{n})`\ricci(\modl_{2})=0;
\otimes_{\struc_{1}}`\natf_{\struc}`\natf_{\conn}`\otimes_{\struc_{2}}]
\efig
\end{equation}

\noindent with the $\natf_{\struc}$ functor above designating a
change in ({\em our} choice of) structure sheaf of generalized
coordinates---from, say, $\struc_{1}$ that might have been chosen
initially, to $\struc_{2}$---and, as a result, from vector sheaf
$\modl_{1}$ to $\modl_{2}$. In turn, $\natf_{\struc}$ induces an
`adjoint' functor $\natf_{\conn}$, which takes us from the
$\otimes_{\struc_{1}}$-functorial vacuum Einstein equations
holding on $\modl_{1}$, to similarly
$\otimes_{\struc_{2}}$-functorial VEG equations holding on
$\modl_{2}$. Clearly, the pair $(\natf_{\struc},\natf_{\conn})$ of
adjoint functor-type of maps above leave the ADG-VEG equations
form-invariant, and have been coined in the past `{\em
differential geometric morphisms}' \cite{rap7}.\footnote{The
ADG-theoretic analogue of the fundamental notion of {\em geometric
morphism} in sheaf and topos theory \cite{maclane1}. Moreover,
since, as noted above, from the ADG-theoretic perspective all
differential geometry is based on $\struc$, differential geometric
morphisms may be equivalently called {\em $\struc$-geometric
morphisms}.} Thus, the most general  ADG-theoretic expression of
the PGC of GR is the following:

\begin{quotation}
\noindent The VEG ADG-field equations (\ref{eq1}) are left
invariant under $\struc$-geometric morphisms.
\end{quotation}

Furthermore, on the PARD and the most general ADG-expression of
the PGC above rests the following {\em Principle of ADG-Field
Realism}(PFR) \cite{rap5,malrap4,rap7,rap11}:

\begin{quotation}
\noindent No matter what $\struc$ we use to (differential
geometrically) represent the gravitational field
dynamics,\footnote{Effectively, {\em our} choice of representation
sheaf $\modl$ on which the connection field $\conn$ acts.} the
latter remains unaffected (:`unperturbed') by our (generalized)
`coordinate measurements' (in $\struc$). The connection
\texttt{field proper} $\conn$ exists `out there', independently of
our coordinatizations (and differential geometric representations)
by $\struc$ (and, {\it in extenso}, by $\modl$ which is locally of
the form $\struc^{n}$). Moreover, since, as noted earlier, if
there is any `spacetime' in our theory, it is encoded in $\struc$,
the ADG-gravitational field $\conn$ and the VEG law (\ref{eq1})
that it defines differential geometrically as a differential
equation proper, `sees through' ({\it ie}, it remains unaffected
by) it. The field $\conn$, and the law (\ref{eq1}) that it
defines, is oblivious to our own coordinatizations/coordinate
measurements in $\struc$ (:$\struc$-functoriality) and therefore
also to the `spacetime geometry' encoded in the latter (by
Gel'fand duality).
\end{quotation}

\item In various, both canonical and covariant, gravitational
field quantization scenaria, the background $M$ and its
$\mathrm{Diff}(M)$ structure (:`symmetry') group create  some
formidable problems. Consider for example the situation in
canonical QGR approached via, say, LQG: there, since the
background spacetime continuum is retained, one regards gravity as
an external spacetime constrained gauge theory; hence, one has to
account for the spatial and temporal diffeomorphism constraints in
the quantum theory (:primary constraints in a Dirac-type of
quantization). This results in notorious problems  that CQGR has
to resolve in order to proceed, such as the problem of defining
meaningful gravitational $\mathrm{Diff}(M)$-invariant quantum
observables,\footnote{Especially in vacuum Einstein gravity
\cite{torre1,torre2}.} the associated problem of finding the
physical Hilbert space of states and its inner product, as well as
the (in)famous problem of time \cite{ish2,kuchar,torre2}. Let
alone that, on top of all this, one has to try to preserve
manifest (general) covariance in the quantum theory when {\it ab
initio} the canonical formalism appears to mandate a $3+1$
space-time split (:a foliation of the base manifold into
space-like hypersurfaces) in order for it to make any sense at all
({\it ie}, to be able to make sense of canonical, `equal-time'
commutation relations between canonically conjugate gravitational
field variables).

Covariant (:path-integral) quantization of gravity scenaria also
encounter challenging obstacles due to the presence of the
background manifold and its $\mathrm{Diff}(M)$ structure group.
The Ashtekar new connection variables 1st -order formalism, for
example \cite{ash}, significantly simplified the constraints in
CQGR and revealed the `innate' gauge-theoretic character of
gravity; albeit, by retaining a base smooth manifold. In
particular, it showed us that the physical configuration space in
the theory is the $\mathrm{Diff}(M)$-moduli space of (gauge)
equivalent spin-Lorentzian connections. It follows that a possible
quantization scenario for gravity could involve a functional
integral over the said moduli space. Thus, an integro-differential
calculus on the aforesaid affine space of smooth connections on a
manifold should be developed \cite{ashlew2,ashlew5}, and the
ever-presence of the infinite-dimensional $\mathrm{Diff}(M)$ group
on the background does not make life any easier. In particular,
one should search for $\mathrm{Diff}(M)$-invariant
Faddeev-Popov-type of  measures on the moduli space of gauge
equivalent connections in order to implement the said functional
integral---a daunting task indeed \cite{baez1,baez2,baez3}.

\end{enumerate}

\noindent At the end of the last section we are going to return
briefly to these QG issues and discuss briefly our ADG-stance
against them. Now however, we would like to go back and dwell a
bit on the issue of functoriality {\it vis-\`a-vis} pre-, 1st-,
2nd- and 3rd-quantization.

Following (the intro to) \cite{mall6}, we note that in much the
same way that the principal aim of geometric (pre)quantization and
2nd-quantization is to bypass 1st-quantization and give directly a
quantum description of relativistic fields ({\it ie}, without
needing first to quantize the corresponding classical mechanical
particle or field theory),\footnote{In this respect, we may recall
from \cite{mall6} the following two contrasting quotes found in
\cite{wood} and \cite{goldstein}, respectively: ``{\em to find a
quantum model of ... an elementary relativistic particle it is
unnecessary ... to quantize [first] the corresponding classical
system}'' and ``{\em ... to quantize a field, we have first to
describe it in the language of mechanics}''. In ADG-field theory,
where, following Einstein \cite{einst3}, ``{\small [In the theory
of relativity,] \em the field is an independent, not further
reducible fundamental concept}...[so that] {\em the theory
presupposes the independence of the field concept}'', it is plain
that we `take sides' with the geometric quantization camp (see
also \cite{zeh} for more, but slightly differently motivated,
arguments against 1st-quantization of a classical mechanical/field
theory).} one can regard as the principal reason for the ADG-based
3rd-quantization as a need for

\begin{quotation}

\noindent{\em a direct quantum description of the ADG-fields `in
themselves',\footnote{This is an autonomous, self-quantization in
accord with the quantum self-duality of the ADG-fields we saw
earlier in this section \cite{malrap4}.}} without any reference to
or dependence on an external (:background to the fields) spacetime
manifold.

\end{quotation}

\noindent Since both the usual geometric (pre)quantization and
second quantization scenaria are essentially rooted on a base
spacetime manifold for their differential geometric formulation in
terms of CDG \cite{brylinski,wood,bandy,echeverria,isidro}, the
ADG-based 3rd-quantization may be seen as an extension and
generalization of both,\footnote{In line with the general fact
that ADG is a significant abstraction and generalization of CDG.}
hence the epithet `third' in order to distinguish between them at
least nominally. An important consequence of this is that while it
is meaningful in 2nd-quantized GR ({\it eg}, QGR approached via
LQG, which is based on the manifold dependent Ashtekar formalism)
to talk about `{\em spacetime continuum
quantization}',\footnote{Something that, as noted earlier, is of
great import in avoiding/resolving gravitational spacetime
singularities in LQG for example \cite{modesto,husain}.} in
3rd-quantized ADG-gravity it is simply meaningless, since no
spacetime manifold, external to the ADG-gravitational field, is
involved ({\it ie}, {\it a priori} assumed in the
theory).\footnote{For instance, as noted before, the evasion of
gravitational singularities in ADG-gravity is secured by the
$\struc$-{\em functoriality} of the ADG-gravitational dynamics
(\ref{eq1}) as singularities of all kinds (even dense, non-linear
distributional ones never encountered before in the mathematics or
physics literature!) are absorbed in $\struc$
\cite{malros1,malros2,malros3,rap5,malrap4}.}

On the other hand, it is well known that geometric prequantization
and second quantization are manifestly functorial procedures. In
what follows we would like to ponder a bit on the functoriality of
3rd-quantization and its physical significance in addition to our
comments on $\struc$-functoriality {\it vis-\`a-vis} gravitational
singularities and QG issues above.

\subsubsection*{Half, first and second quantization} Broadly speaking,
{\em prequantization}, or what we here coin `{\em half
quantization}', pertains to a formal mathematical procedure which
establishes a {\em correspondence} between the classical
description of a physical system and a quantum description of the
same system. Given the standard mathematical model of the
kinematical space of a classical mechanical system as a smooth
phase space $M$ (:differential manifold) endowed with a symplectic
structure $\omega$ on it (:a symplectic manifold, write
$S=(M,\omega)$), together with a Hamiltonian function $H$
generating its smooth dynamical (:time) evolution,
$\frac{1}{2}$-quantization corresponds it to a Hilbert space
$\mathcal{H}$ in such a way that the transformations of $M$
preserving $\omega$ (:the canonical or symplectic maps) are mapped
to unitary operators on $\mathcal{H}$, which by definition
preserve $\mathcal{H}$'s inner product
(:isometries).\footnote{Usually one assumes $\mathcal{H}$ to be
$L^{2}(M)$---the Hilbert space of smooth ($\com$-valued) square
integrable functions on $M$ relative to the standard Liouville
measure $\mu_{M}$ on the latter, defining a Hermitian inner
product: $< \phi | \psi>:=\int \phi^{*}\psi d\mu_{M}$, with
$\phi^{*}$ the complex conjugate of $\phi$.} In {\em
category-theoretic} terms,

\begin{quotation}

\noindent{\em prequantization is a functor from the category of
symplectomanifolds and canonical morphisms, to the category of
Hilbert spaces and unitary operators on them.}

\end{quotation}

On the other hand, it is also well known that {\em first
quantization}, unlike prequantization, is {\em not} a functorial
procedure. By 1st-quantization one means a correspondence, like
prequantization, between the aforesaid symplectic and Hilbert
space categories which furthermore carries a one-parameter group
of canonical transformations generated by a {\em positive} $H$, to
a one-parameter group of unitaries generated by a {\em positive}
Hamiltonian operator $\hat{H}$. The bottom line here is that

\begin{quotation}
\noindent {\em first quantization is not  functorial}, because
energy-positivity is not preserved in transit from the classical
to the quantum description.
\end{quotation}

However, if one has established a single-quantum (:particle)
description of a physical system, one can pass {\em functorially}
to a many-particle one ({\it eg}, a quantum field) by the process
of {\em second quantization}. Briefly, starting from the
single-quantum Hilbert space $\mathcal{H}$ above, and depending on
the spin of the particle-quanta of the fields considered, one
takes completely symmetric or antisymmetric tensor products of any
number of identical copies of $\mathcal{H}$, which when directly
summed and completed yields the so-called Fock state space in
which free quanta of the corresponding boson and fermion fields
are supposed to live. Thus, the {\em 2nd-quantization functor} is
from the Hilbert category to itself (appropriately
tensored),\footnote{That is, the Hilbert category is augmented by
the usual tensor product $\otimes$, hence it is regarded as a
tensor monoidal category.} and it can be easily seen to be
positivity preserving.

To summarize, while prequantization and second quantization are
functorial constructions (procedures or correspondences), first
quantization is not. Actually, as noted above, it is the {\it
raison d'$\hat{e}$tre et de faire} of {\em geometric
(pre)quantization} to bypass completely 1st-quantization and
describe directly 2nd-quantized (:quantum) fields, without
recourse to a classical mechanical particle or field system which
has to be quantized first.\footnote{Indeed, on general
philosophical grounds, it is {\em unnatural} to suppose that there
is a classical/quantum dichotomy in Nature (pun intended). Quantum
theory is the fundamental description, while all classical ones
are coarse approximations (:effective descriptions) of the
fundamental quantum one \cite{malrap4,rap5,rap11}. To paraphrase
Finkelstein: ``{\em all is quantum; anything that appears to be
classical has not been resolved yet into its quantum elements}''
\cite{df2}.}

\subsubsection*{ADG in  second and geometric (pre)quantization: further general remarks}

The aforesaid bypass of 1st-quantization by geometric quantization
in order to arrive directly at a quantum description of fields
suits perfectly ADG, since in the latter the basic objects---its
fundamental building blocks or {\it ur}-elements so to speak---are
ADG-fields of the $(\modl ,\conn)$ kind. Thus, in ADG-field
theory, in keeping with the terminology and technical machinery of
the manifold and CDG-based geometric (pre)quantization
\cite{brylinski,wood,bandy,echeverria,isidro}, albeit in a
manifestly background manifold independent way as befits ADG, the
epithet `geometric' to `(pre)quantization' pertains to the
identification of the $\omega$ of a symplectic manifold---a closed
differential 2-form on $M$---to the curvature $\curv(\conn)$ of a
connection field $\conn$ on a (Hermitian) Hilbertian
representation vector sheaf $\modl$. The latter can be regarded,
via an ADG-theoretic generalization  of the celebrated  Serre-Swan
theorem \cite{mall6,mall4} originally motivated by some arguments
of Selesnick in the context of 2nd-quantization \cite{sel}, as a
{\em free, finitely generated projective $\struc$-module, whose
(Hilbert space) stalks represent the (pre)quantum state spaces of
the ADG-field systems in focus} \cite{mall5,mall6,mall4}. In turn,
(the curvature of the connction on) $\modl$ obeys some sort of
{\em quantization condition} ({\it eg}, Weil's integrality
condition), which is instrumental in classifying sheaf
cohomologically the vector sheaves involved ({\it eg}, Chern-Weil
theorem, Chern characteristic classes, the Picard group)
\cite{mall2,mall5,mall4}. Moreover, as we highlighted it earlier
in section 2, from a 2nd-quantization vantage the local sections
of the Hilbertian $\modl$s represent {\em local quantum particle
states of the corresponding fields}, while also an ADG-theoretic
version of the spin-statistics connection comes to identify {\em
local free boson states with local sections of line sheaves},
while {\em local particle states of fermionic fields correspond to
local sections of Grassmannian vector sheaves of finite rank
greater than 1}.

All this is well done and dusted; however, here we would like to
make a couple of additional scholia in the light of
3rd-quantization presently proposed and its
$\struc$-functoriality:

\begin{enumerate}

\item Quantum fields are traditionally regarded as special
relativistic entities,\footnote{After all, QFT is supposed to be a
unison of SR and QM (:quantum fields over flat Minkowski space).}
hence their quantum particle states have been modelled after
irreducible representations of the Poincar\'e group {\it \`a-la}
Wigner. At the same time, what has for many years stymied efforts
to genuinely unite relativity with quantum theory in a finitistic
setting is the fact that, because the Lorentz group is
non-compact, there are no {\em finite-dimensional} unitary
representations of it \cite{haag}.\footnote{In \cite{df0} for
example, this shortcoming was diagnosed early in building the
`Space-Time Code', or subsequently in developing the quantum net
dynamics \cite{df1,df3}, and Finkelstein opted for sacrificing
quantum mechanical unitarity (but preserve algebraic finiteness!),
because it is a non-local notion (as it involves an {\em integral}
over all space). Furthermore, for quantum fields over a curved
spacetime manifold (in case for instance one wished to apply the
QFTheoretic formalism `blindfoldedly' to gravity in a
semi-classical manner) the situation is even worse, since there
are inequivalent (unitary) representations of the gauged
(:spacetime manifold localized) Lorentz group \cite{fulling}.}
Yet, the reader must have already observed that our representation
vector (:Hilbert $\struc$-module) sheaves $\modl$ are of finite
rank, and they constitute unitary representation spaces of the
`internal' (:gauge) symmetry groups, which are of course compact
\cite{mall6,mall4}. There is no discrepancy here: 3rd-quantum
field theory does not involve any base spacetime at all---be it
flat Minkowski space or a curved gravitational background; hence,
we do not have to account for a potential conflict between
finite-dimensionality of (unitary) particle representations and
the Lorentz group.\footnote{In any case, it has been argued for a
long time now whether exact (local) Lorentz invariance would
survive in the finitistic QG domain (:below its Planck length
`cut-off'). Especially in the supposedly inherently discrete
setting of Sorkin's causal set theory \cite{,sork1,sork2}, the
issue of whether Lorentz invariance should be preserved or given
up for good has recently become a caustic one, with current
tendencies leaning towards abandoning it
\cite{henson1,henson2,henson3}.} {\it In summa}, {\em no
background spacetime, no external spacetime symmetries; and all
the symmetries of the ADG-fields are `internal'
(:gauge)}---another positive feature of the 3rd-gauge 3rd-quantum
ADG-fields' autonomy.

\item As noted earlier, the central result of geometric
(pre)quantization is the identification of the symplectic form
$\omega$ on the $\smooth$-manifold phase space of a physical
system with the curvature of a connection (on the same manifold).
The aforementioned $\struc$-functoriality of the ADG-VEG dynamics
(\ref{eq1}) comes in handy here since the latter is an expression
involving the geometrically (pre)quantized curvature $\curv$ of
the ADG-gravitational \texttt{field proper} $\conn$, and
$\curv(\conn)$ is an $\otimes_{\struc}$-tensor (:an
$\struc$-morphism). Precisely in this sense we maintained in
section 2 that we already possess a geometrically
(pre)quantized---and, in view of \cite{mall6}, a
2nd-quantized---vacuum gravitational (and, {\it in extenso}, free
YM) dynamics. Moreover, this dynamics is manifestly generally
covariant (in the generalized ADG-sense of general covariance
involving $\aut\modl$), and it is of course also left
form-invariant under $\struc$-geometric morphisms. Finally, it is
straightforward to see that the sheaf cohomological local
Heisenberg uncertainty relations (\ref{eq9}) defining
3rd-quantization remain invariant under $\struc$-geometric
morphisms, something that further vindicates the `universality'
(:`functoriality') of 3rd-quantization.

\item The final remark we wish to make here is a twofold, quite
general one, bearing historical/methodological undertones and
going at the heart of ADG {\it vis-\`a-vis} applications of
differential geometry to the quantum (gauge field) domain. First,
we must highlight the use of {\em sheaves} (and sheaf cohomology)
instead of {\em fiber bundles} in gauge field theory that ADG
advocates. Fiber bundles is {\em the} mathematical theory
currently used in (applications of differential geometric ideas to
quantum) gauge (field) theories. However, it has been recently
felt that fiber bundles come short in modelling what's `really'
happening in the quantum field and, {\it in extenso}, in the QG
regime, and should thus be replaced by  sheaves. We let Haag
\cite{haag} and Stachel \cite{stachel3,stachel2} do the talking
here:

\begin{quotation}
\noindent ``{\small ...{\bf Germs.}  We may take it as the central
message of Quantum Field Theory that all information
characterizing the theory is strictly local i.e. expressed in the
structure of the theory in an arbitrarily small neighborhood of a
point.\footnote{Our emphasis.} For instance in the traditional
approach the theory is characterized by a Lagrangean density.
Since the quantities associated with a point are very singular
objects, it is advisable to consider neighborhoods. This means
that instead of a fiber bundle one has to work with a sheaf. The
needed information consists then of two parts: first the
description of the germs, secondly the rules for joining the germs
to obtain the theory in a finite region...}''

\vskip 0.1in

\noindent \underline{\bf and}

\vskip 0.1in

\noindent ``{\small ...However, the fibre bundle approach clearly
does not solve the second problem discussed in the previous
section.\footnote{That is, not fixing up-front the global topology
of the manifold, and globalizing a local solution to the Einstein
equations---{\it in toto}, (analytically) extending a local
solution (a local region where the law holds) to a global one (the
total spacetime manifold where the law is valid).} The topology of
the base manifold is given {\it a priori}, so that a different
fibre bundle must be introduced {\it a posteriori} for each
topologically distinct class of solutions. The process of finding
the global topology cannot be formalized within the fibre bundle
approach. It appears that sheaf cohomology theory is the
appropriate mathematical theory for dealing with the problem of
going from local to global solutions...}'' \cite{stachel3}

\vskip 0.1in

\noindent \underline{\bf also}

\vskip 0.1in

\noindent ``{\small ... sheaf theory might be the appropriate
mathematical tool to handle the problem\footnote{The problem noted
above: going ({it eg}, extending a solution to the field
equations) from `local' to `global'.} in general relativity. As
far as I know, no one has followed up on this suggestion, and my
own recent efforts have been stymied by the circumstance that all
treatments of sheaf theory that I know assume an underlying
manifold...}'' \cite{stachel2}

\vskip 0.1in

\end{quotation}

\noindent  Indeed, a virtue of sheaf theory is that it `naturally'
effectuates easily  (virtually by the very definition of a sheaf)
the much desired transition from local to global (or `micro' to
`macro'), and by its very definition models (dynamically)
`variable structures'.\footnote{Briefly, {\em localization (of a
structure) is gauging it, and gauging (a structure) is tantamount
to making it (dynamically) variable}---{\it ie}, endowing it with
a dynamical connection field.} We should also stress here that
sheaf theory, at least  as it has been developed and used in ADG,
does not involve at all any base manifold like the sheaves in
Minkowski space (QFT) or over a general curved spacetime manifold
(QG) that Haag (implicitly) and Stachel (explicitly) allude to in
the quotations above.

It is about time theoretical physicists (in prticular, quantum
field theorists)  broke from the mold of bundles and became
acquainted with the rudiments of sheaf theory. Yet, one has to
appreciate the hitherto unprecedented in the mathematics
literature use of sheaf theory, in all its generality, power and
resourcefulness, in {\em differential} geometry that ADG has
brought about That one can do differential geometry purely
algebraically,  independently of any notion of `smoothness' in the
standard sense (of employing a background $\smooth$-manifold to
`mediate' our differential calculus) is indeed a feat of ADG that
could have enormous implications (and applications) in current QG
and quantum gauge theory research.

The second ptyche of ADG we would like to highlight is the use of
general, possibly  {\em non-normed}, topological algebras in the
quantum (field) regime. Briefly, ever since the von Neumann
quantum axiomatics in Hilbert space, quantum (field) theorists
have (pre)occupied themselves with the study of von Neumann and
$C^{*}$-algebras, the `canonical' example being the non-
$C^{*}$-algebra  ${\mathcal{B}}(\hil)$ of bounded operators on
Hilbert space after which algebras of quantum mechanical
observables are usually modelled \cite{bratteli,haag}. On the
other hand, the archetypical example of a non-normable
(:non-Banachable) topological algebra is $\smooth(M)$---in fact,
the {\em only} algebra we have used so far to do {\em
differential} geometry (:CDG on manifolds). Admittedly, one could
try to retain non- $C^{*}$-ness and try to develop a differential
geometry based on such algebras, as in the `Noncommutative
Calculus' of Connes \cite{connes}. Yet, one could object even `in
principle' to such an endeavor by observing that operators of
quantum physical interest such as `position' are essentially {\em
unbounded}, while at the same time, the reason behind the use of
commutative algebras as structure sheaf of generalized coordinates
(as it is the case in ADG) is Bohr's correspondence principle.
Namely that, while quantum mechanical actions are noncommutative,
{\em our} measurements (ultimately, our geometrical
representations and interpretations!) of them are essentially
commutative.\footnote{As mentioned earlier, at the bottom even of
Connes' functional-analytic/operator-theoretic noncommutative
geometry lies the manifold $M$, with its commutative coordinates.}
The bottom line here is, to paraphrase Borchers this time, that
physicists should break free from Banach algebras (essentially,
from the mold of Euclidean spaces, finite or
infinite-dimensional!) and familiarize themselves with the theory
of topological algebras (especially non-normed ones), which may be
of great import in many physical applications
\cite{naimark,kolmogorov}. .

\end{enumerate}

\noindent Now, in the next paragraph we vindicate some of the
remarks above in the light of \cite{mall6}. In particular, based
on Mallios' `universal' $K$-theoretic musings on  2nd-quantization
under the prism of ADG as exposed in that paper, we draw some
telling links with our 3rd-quantization of VEG and FYM theories.
In addition, we make contact with our (f)initary, ©ausal and
(q)uantal (:$fcq$) VEG developed in the hexalogy
\cite{malrap1,malrap2,malrap3,rap5,malrap4,rap7}.

\paragraph{$\mathbf{K}$-theoretic underpinnings of 3rd-quantization and a link with $fcq$ ADG-VEG.}
We can relate the heuristic canonical-type of 3rd-quantization
introduced above with Mallios'  $K$-theoretic musings on
2nd-quantization {\it \`a la} ADG in \cite{mall6,mall4}.

In \cite{mall6}, cogent arguments are given for representing the
quantum state spaces of (free) elementary particles of quantum
(:2nd-quantized) fields by the vector sheaves (:locally free
$\struc$-modules of finite rank) involved in ADG. The syllogism
supporting those arguments takes us progressively from {\em free}
(Hilbert) {\em $\struc$-modules}, to {\em finitely generated free
$\struc$-modules}, to {\em finitely generated projective
$\struc$-modules}, and finally, via an extension of the classical
Serre-Swan theorem to non-normed topological
algebras,\footnote{The classical Serre-Swan theorem bijectively
corresponds finitely generated projective $\cont(X)$-modules (with
$X$ a compact topological manifold, and $\cont(X)$ the algebra of
continuous $\com$-valued functions on it) to continuous complex
vector bundles over $X$. Moreover there are smooth analogues of
that correspondence, in that one can map bijectively
$\smooth(X)$-modules (with $X$ now a compact $\smooth$-smooth
manifold) to $\smooth$-vector bundles over it. Mallios' extension
of the classical result consists in allowing for more general
(than $\smooth(X)$) non-normed topological algebras  for
coefficient algebras.} to vector bundles and their associated
vector sheaves $\modl$ that ADG is all about.\footnote{The
original motivation for this syllogism was Selesnick's paper
\cite{sel}. More technical details of the arguments and their
associated constructions, of the closely related ADG-version of
the spin-statistics connection based on what is there coined {\em
Selesnick's Correspondence}, as well as the various relevant
references backing those arguments can be found in
\cite{mall6,mall4}.}

Regarding our brief remarks at the end of the last paragraph about
the use of sheaves instead of bundles and of general (:non-normed)
topological algebras instead of Banach ones, two points in
\cite{mall6} must be highlighted here, namely:

\begin{enumerate}

\item That once one arrives by the above syllogism and its related
constructions at vector {\em sheaves} as a model for the state
spaces of field-quanta, one forgets altogether about bundles and
works exclusively with the (local) sections of the resulting
sheaves;

\item That of special interest (and use!) are certain `nice'
non-normed commutative topological algebras
$\struc$,\footnote{Locally $m$-convex $Q$-algebras ({\it alias},
Waelbroeck algebras) \cite{mall0}. An archetypical example of a
Waelbroeck algebra is $\smooth(X)$, with $X$ a compact Hausdorff
$\smooth$-manifold. Note that here we abuse notation and use also
$\struc$ for the said algebras (when Mallios uses $\mathbb{A}$).
We hope that the reader will not confuse the `algebras' with the
`structure sheaves' thereof, but anyway the distinction is going
to be clear from the context.} which are seen to be localized
sheaf-theoretically over their {\em Gel'fand spectra}
$\gelsp(\struc)$ \cite{mall-1}, which in turn are `topologically
indistinguishable' ({\it eg}, {\em homotopic}) to the base
topological space $X$ over which the $\struc$-module sheaves were
defined in the first place. This is a nice example of {\em
Gel'fand duality} and it highlights what we emphasized earlier:
that if any `space(time)' is involved at all in our scheme, it is
already encoded in $\struc$, while at the same time one works
solely in `sheaf space' $\modl$ ({\it ie}, with $\modl$'s
sections) with a purely algebraic (:sheaf-theoretic) `differential
geometric mechanism' that is manifestly $\struc$-functorial. Thus
one essentially forgets about `space(time)' altogether.

\end{enumerate}

\noindent Keeping in mind points 1 and 2 above, in \cite{mall6} an
elegant $K$-theoretic formulation and of the Serre-Swan theorem
(extended to non-normable topological algebras) is given involving
{\em Grothendieck $K$-groups}.\footnote{See \cite{mall6} for
details of the constructions, and \cite{mall0} for the technical
terminology and notation used therein.} In a nutshell, modulo an
group isomorphism, one establishes the following equalities

\begin{equation}\label{eq20}
K(X)=K(\struc)=K(\mathcal{P}(\struc))
\end{equation}

\noindent so that {\em $X$ is homotopic to $\gelsp(\struc)$},
while the latter is assumed to be a unital, commutative,
associative, locally $m$-convex $Q$-algebra (:Waelbroeck algebra).

What is of interest to us here {\it vis-\`a-vis} 3rd-quantization
is that in \cite{mall6} (\ref{eq20}) {\em is further expressed in
terms of projection operators}. Briefly, one singles out {\em a
primitive idempotent linear endomorphism} $\proj$

\begin{equation}\label{eq21}
\proj\in M_{n}(\struc):~\proj^{2}=\proj
\end{equation}

\noindent whose {\em kernel} $\mathrm{ker}(\proj)$ corresponds to
a {\em finitely generated projective $\struc$-module
$\mathcal{M}$}

\begin{equation}\label{eq22}
\mathcal{M}=\mathrm{ker}(\proj)
\end{equation}

\noindent defining a Grothendieck class $[{\mathcal{M}}]$ (of
finitely generated projective $\struc$-modules) in $K(\struc)$. In
turn, $\proj$ is seen to lift to a {\em morphism}

\begin{equation}\label{eq23}
\hat{\proj}:~X\times\struc^{n}\rightarrow X\times\struc^{n}
\end{equation}

\noindent whose kernel,
$[\mathcal{M}]=[\mathrm{ker}(\hat{\proj})]$, is now a Grothendieck
class of modules in $K(X)$. Moreover, one easily observes that
$\mathrm{ker}(\proj)$ defines a continuous finite-dimensional
$\com$-vector bundle $(E,\pi ,X)$ over $X$, from (the sections of)
which one arrives straightforwardly to a vector sheaf $\modl$ as
defined in ADG.

Regarding our 3rd-quantization scenario of ADG-fields ({\it ie},
vector sheaves carrying connections), the alert reader may have
already spotted the `caveat' in the $K$-theoretic construction
above:

\begin{quotation}
\noindent {\em The projection operator $\proj$ in} (\ref{eq21})
{\em may be identified with our sheaf cohomological quantum
uncertainty operator} $\unc$ {\em in} (\ref{eq13}).
\end{quotation}

\noindent The full physical meaning and implications of this
identification ought to be further explored and better
comprehended \cite{rap13}.

Finally, since the present paper is a contribution to Sorkin's
60th birthday fest-volume, it is fitting to make some comments on
Mallios' generalization of the preceding $K$-theoretic musings
(:expressions (5.12) and (5.13) in \cite{mall6}) in the light of
our applications of ADG to `finitary, causal and quantal' VEG. In
the said expressions, Mallios establishes that

\begin{equation}\label{eq23}
K(X)=K(\struc)
\end{equation}

\noindent when  $X$ is the projective limit of an inverse system
$\underleftarrow{\{ X_{i}\}}$ of topological spaces
$(:X=\underleftarrow{\lim} X_{i})$, each of which is the Gel'fand
spectrum of a Waelbroeck algebra $(:X_{i}=\gelsp(\struc_{i}))$,
with the latter constituting an inductive system
$\underrightarrow{\{ \struc_{i}\}}$ whose direct limit is $\struc$
itself $(:\struc=\underrightarrow{\lim}\struc_{i})$.

The reader who is familiar with our `finitary' work
\cite{rapzap1,rapzap2,rap1,rap2,malrap1,malrap2,malrap3,rap5,malrap4,rap7},
can easily translate the $K$-theoretic result above to the
finitary case where $\underleftarrow{\{ X_{i}\}}$ is taken to be
the projective system of Sorkin's finitary poset substitutes
$P_{i}$ of (a relatively compact region $X$ of) a topological
manifold $M$ \cite{sork0} and $\underrightarrow{\{ \struc_{i}\}}$
an inductive system of incidence (Rota) algebras $\Omg_{i}$, whose
primitive Gel'fand spectra are precisely the $P_{i}$s
($\gelsp(\Omg_{i})=P_{i}$).\footnote{Discrete Gel'fand duality
between finitary posets and their incidence algebras
\cite{zap0,rapzap1,rapzap2,rap1}.} This ($K$-)functorial
correspondence has been anticipated in \cite{zap1,zap2}, and the
direct/inverse limits engaged in it have been interpreted as
`classical limits' \cite{rapzap1,rapzap2}. These observations too
must await a more thorough investigation \cite{rap13}.

\paragraph{A final note on terminology.} In closing this paper we would like
to mention parenthetically that the name `{\em third
quantization}' has already been used in the theoretical physics
nomenclature, pertaining to some ideas in early universe cosmology
\cite{strominger}. We should emphasize that our 3rd-quantization
has little in common with that original term, so that the two
should not be confused or thought to be somehow related. Also, the
same term has been used by John Baez in his general theoretical
scheme that might be coined `higher-order categorical
quantization', {\it alias}, `$n$-th quantization' ($n\geq2$) for
short \cite{baez0}. Here again, apart from the common formal
algebraico-categorical language and technology underlying both our
3rd-quantization and Baez's, there is little common semantic
grounds between the two schemes.

\section{Summary with Concluding Remarks}

In the present paper we gathered certain central results from
manifold applications of ADG to gravity and gauge theories and
argued that we already possess a geometrically (pre)quantized,
second quantized and manifestly background spacetime manifold
independent vacuum Einstein gravitational and free Yang-Mills
field dynamics. Based on the {\it ur} ADG-conception of a field as
a pair $(\modl ,\conn)$, we entertained the idea of a field
quantization scenario called third quantization. 3rd-quantization,
like geometric (pre)quantization and second quantization, was seen
to be an expressly functorial scheme which, in contradistinction
to its two predecessors, does not depend at all on a background
manifold for its differential geometric formulation and physical
(:spacetime continuum) interpretation. It thus extends them both,
following the extension and generalization of the Classical
Differential Geometry (CDG) of $\smooth$-smooth manifolds that ADG
has achieved by purely algebraico-categorical (:sheaf-theoetic)
and sheaf cohomological means.

In what formally looked like canonical quantization, but in the
manifest absence of a smooth background geometrical spacetime
manifold as befits ADG, we posited abstract non-trivial local
Heisenberg-like commutation relations between certain
characteristic local (:differential) forms that uniquely
characterize sheaf cohomologically the ADG-fields and their
particle-quanta. These forms were then physically interpreted in a
heuristic way as abstract position and momentum `determinations'
(:`observables') in accordance with ADG's (pre)quantum field
semantics. The ADG-fields were thus said to be `third quantized',
and so are the vacuum Einstein and free Yang-Mills equations that
they define differential geometrically as differential equations
proper, without any need arising to quantize an (external to the
ADG-fields) spacetime continuum, simply because such a theoretical
artifact does not exist in our theory. Due to the explicit
functoriality of our ADG-constructions, as well
 as the background spacetime manifoldlessness that goes hand in hand with it,
 3rd-quantization was seen to be fully covariant and it totally bypasses second
 quantized gravity's vital reliance on a base $M$ and its diffeomorphism structure
 group $\mathrm{Diff}(M)$ for its differential geometric formulation and physical
 interpretation as an external spacetime continuum constrained quantum gauge theory.
 All in all, we maintained that

\begin{quotation}
\noindent {\em ADG-VEG is a purely gauge, external spacetime
manifold unconstrained, third quantized theory.}
\end{quotation}

Third quantized ADG-gravity's full covariance and background
manifoldlessness  motivates us to view certain outstanding and
obstinately resisting (re)solution current QG problems under a new
light. Thus, as a future project we entertain the possibility of
developing a genuinely covariant functional integral quantization
of vacuum Einstein gravity (and free Yang-Mills theories). The
functional integral will be over the moduli space of
$\aut\modl$-equivalent $\struc$-connections, which is the physical
configuration space in ADG-gravity.  A generalized Radon-type of
measure, as recently developed in \cite{mall4}, will be used to
implement the functional integral. What is more important,
however, is to note that, due to the manifest background spacetime
manifoldlessness of ADG-gravity, we expect such an abstract path
integral scenario to be free from the problem of finding
$\mathrm{Diff}(M)$-invariant measures on the moduli space of gauge
equivalent connections, which has so far stymied the usual
manifold and CDG-based path integral approaches. In this way, we
will catch glimpses of a genuine equivalence between the `local'
(:differential) canonical-type 3rd-quantized gravity and a
potential `global' functional integral-type of one. Of course, the
methods of sheaf theory, especially as they have been developed
and used by ADG, enable us to address both local (:differential
canonical) and golobal (path integral) quantization issues. In any
case, such an equivalence is formally absent from the usual base
spacetime manifold and CDG-based quantization approaches, since,
for instance, the smooth base spacetime dependent canonical
quantum gravity manifestly breaks covariance in two ways. On the
one hand, it mandates a $3+1$ space-time split (:foliation of
spacetime into spacelike hypersurfaces) in order to concord with a
well posed Cauchy problem in the classical theory (:GR) that it
purports to quantize, while on the other, in order to adapt
consistently the usual canonical formalism and its physical
interpretation to the said foliation, it posits equal-time
commutation relations between the conjugate gravitational field
variables restricted on the aforesaid spatial hypersurfaces. Even
in the usual supposedly covariant (:path integral) quantization
schemes, whether Lorentzian or Euclidean (which apparently does
not formally distinguish between space and time `directions' {\it
ab initio}), input and output field amplitude data still have to
be specified on initial and final hypersurfaces respectively in
order to have a meaningful path integral quantum dynamical
propagator between them.

Finally, it is plain that the manifest absence of a background
spacetime manifold in 3rd-quantized ADG-VEG prompts us to
emphasize that our scheme evades totally the {\em problem of time}
\cite{ish2,kuchar}, the {\em inner product problem}, as well as
the {\em problem of defining meaningful gravitational observables
in VEG} \cite{torre1,torre2}, all of which, in one way or another,
hinge on our regarding $\mathrm{Diff}(M)$ automorphism `structure'
group of the underlying $\smooth$-smooth manifold of GR as
gauge-constraining the gravitational field, by implementing the
PGC.

\paragraph{Declaration.} The intuitive and heuristic ideas presented here are `raw' and under development, hence they must await a
more formal treatment and a mathematically rigorous exposition
\cite{rap13}.

\section*{Acknowledgments}

It is a pleasure to acknowledge numerous inspiring interactions
during a long collaboration with Tasos Mallios on potential
applications of ADG to quantum gravity and gauge theories. This
work has been financially supported by the European Commission in
the form of a generous Marie Curie European Reintegration Grant
(ERG-CT-505432) held at the Algebra and Geometry Section,
Department of Mathematics, University of Athens (Greece). Material
support from the Theoretical Physics Group at Imperial College
(London) is also greatly appreciated. Finally, I wish to thank Fay
Dowker for giving me the opportunity to contribute a paper to this
`{\em Sorkin Fest}' volume.

\section*{Postscript: Brief encounters of the third kind}

It is with great pleasure that I contribute the present paper in
honour of Rafael Sorkin. In what follows, I would like to sketch
out perigrammatically a short memoir of my crossing worldlines,
interaction vertices and scattering cross-sections with Rafael and
his work. In keeping with the central  notions of gauge theory of
the 3rd kind and 3rd-quantization presented in this paper, the
personal account of my brief `other worldly'  experiences with Ray
and his research may be fittingly coined `{\em brief encounters of
the 3rd kind}'.

\paragraph{Meeting worldlines and interaction vertices.} My interest in
Rafael's work began quite unexpectedly back in 1992 when, as a
first-year doctoral student at the University of Newcastle upon
Tyne, I accidentally bumped into a pre-print of a Greek sounding
physicist---Charilaos Aneziris---titled {\it Topology and
Statistics in Zero Dimensions}. In that paper, some interesting
links between discrete topology and the quantum spin-statistics
connection were drawn. The work was based primarily on a paper by
Rafael Sorkin titled `{\it Finitary Substitute for Continuous
Topology}' (abbr. FSCT hereafter) written a year or so earlier
\cite{sork0}.

I thus tracked down the latter via British Interlibrary Loans (for
the small Armstrong library at Newcastle did not keep an
up-to-date stock of IJTP) and found it profound and masterfully
written. I was particularly impressed by the simplicity of ideas
and the `unassuming' character of Rafael's writing. I was (I guess
I still am) a novice in QG research, thinking that the Holy Grail
of modern theoretical physics would somehow have to involve
intricate physical reasoning, dressed up in a fancy, almost
cryptic, mathematical language. I was dumbfounded to find Rafael's
`finitary stuff' deep, yet simple; original and fresh, yet as if I
had subconsciously already thought about it somehow (or at least,
I felt ready to sit down and do research on it!).\footnote{Indeed,
I later realized that the profundity of Rafael's papers lies in
their laconic, `ostensive', almost `{\it  in-your-face}', style
and subject matter. They lay bare what is at stake and they expose
their subject in the simplest {\em conceptual} language possible,
virtues that I had previously encountered only in (some, mainly
philosophical) post-30s Einstein writings, as well as in (some of)
Feynman's Lectures in Physics.}

In the FSCT, I first came across the causal set (:causet)
scenario. I then read the seminal `{\it Bomb Lee, Me and Sorkin}'
paper \cite{bomb87} and for a while I got hooked on causets, if
only day-dreaming about them. At about the same time, I came
across David Finkelstein's work on the `{\em Space-Time Code}'
\cite{df0} and his `{\em Superconducting Causal Nets}' \cite{df1}
(the second written two decades after the first), in which I found
the primitive seeds for a `{\em quantum algebraization of discrete
causality}'. Then, I recall my first brief  {\it
t\^{e}t-a-t\^{e}t} meeting with Rafael at a coffee break during
the 3rd Quantum Concepts in Space and Time conference, organized
by Chris Isham and Roger Penrose, in Durham (July 1994).  There, I
doubt whether Rafael remembers our fleeting encounter, I recall
pitching to him the idea of algebraizing discrete causality {\it
\`a la} Finkelstein, and of the general idea of finding an
algebraic structure to encode a locally finite poset (be it a
finitary topological substitute of the continuum, or a causet).

Already a decade earlier, however, there was a sea-change in
Rafael's thinking about locally finite partial orders: from their
original inception as coarse topological approximations of the
spacetime continuum, to their being regarded as fundamental
discrete causal structures to which, in the other way round, the
Lorentzian  spacetime manifold of GR is now a coarse
approximation. This reversal in the physical interpretation of
finitary porders is nicely accounted for in
\cite{sork1}.\footnote{Another example of a simple, fluent and
conceptually deep paper.} On the other hand, the FSCT paper made a
deep and lasting impression on me in that the primitive idea was
suggested to replace the operationally ideal and `singular' points
of the point-set spacetime continuum by `fat' regions (:open sets)
about them, the latter belonging to locally finite coverings of
the topological manifold we started with. Then, relative to such
covers, Rafael extracted a finitary poset, which, moreover, had
the structure of a `pointless' simplicial complex. It is not an
exaggeration to say that the said `pointlessness' and
`algebraicity' were two of my original motivations for applying
category and in particular topos-theoretic ideas to QG in my Ph.D.
thesis. Such was Rafael's influence.

Soon after I got my doctorate, I became familiar with Tasos
Mallios' ADG theory, and as a postdoc at the maths department of
the University of Athens I met Roman Zapatrin who gave a most
interesting seminar on a possible algebraization of Rafael's
finitary-topological posets (:fintoposets) using so-called {\em
incidence (Rota) algebras} \cite{zap0}. Roman invited me to
St-Petersburg, where we wrote our first collaborative paper on a
possible algebraic quantization scheme for the fintoposets of the
FSCT paper based on the incidence algebras thereof \cite{rapzap1}.
Shortly after, by analogy to the topological case, but now bearing
in mind the aforesaid semantic reversal in \cite{sork1}, I
conceived of a similar `algebraic quantization' scenario for
causets \cite{rap1} using a discrete version of {\em Gel'fand
duality} originally due to Roman.

Here I shall digress a bit and tell you a telling little anecdote:
in the early summer of the millennium year (:six years after I had
first met Rafael in Durham!), when I was a maths  postdoc at the
University of Pretoria, I e-mailed Rafael, excited about my
algebraic quantization of causets scheme. I never received any
reply from him during the whole summer, thus I was quite
disappointed and thought that my ideas were not that interesting
to causet people after all. However, in mid-September I
unexpectedly received the following laconic, almost telegraphic,
2-line e-message: \footnote{The following is a reconstruction from
memory of Rafael's message, but it is pretty accurate (especially
the last 4-word expression).}

\begin{quotation}
\noindent ``{\em I know of your work. You formulated Gel'fand
duality for causets before Djamel Dou\footnote{A doctoral
student-collaborator of his at Syracuse University back then, I
believe.} and I did, thus there is no need for us to `beat around
the bush'. }
\end{quotation}

\noindent This message highlights nicely the following triptych of
traits of Rafael's character (of course, with a bit of
generalization written in inverted commas below):

\begin{enumerate}

\item He `always' answers laconically and to the point;

\item He `never' answers to his e-mail messages promptly;

\item He is ever ready to acknowledge the work of other people and
to give credit, when credit is due;\footnote{Although in this
case, proper credit should have been given to Roman, for the
Gel'fand duality for causets \cite{rap1} comes  {\it mutatis
mutandis} from the corresponding duality between finitary
topological posets (:simplicial complexes) and their incidence
algebras \cite{zap0, rapzap1,rapzap2,zap1}.}

\item In retrospect, especially after the appearance of his fairly
recent paper \cite{sork2}, it is plain to me that Rafael never
regards a robust and beautiful result, such as discrete  Gel'fand
duality for finitary posets, as `closing the matter' ({\it ie},
that there is no need of `beating around the bush'). For the bush
is always out there to be beaten,\footnote{Part and parcel, I
guess, of Rafael's `hard-core' physico-philosophical realism.} in
the sense that a result can always be improved, hence for instance
his discovery of ideals in incidence algebras better suited to the
causet structure and its physical semantics than our Gel'fand
ideals \cite{sork3}.

\end{enumerate}

\paragraph{Scattering cross-sections and diverging amplitudes.}\footnote{I am quite sure that Rafael,
who is the epitome of modesty, won't feel comfortable with the
verbose eulogy that follows. I apologize in advance, but in a way
I feel `obliged' to do it, plus I don't know of another way of
expressing what I wish to say and what I feel.} I noted earlier my
coming across Mallios' ADG in the late 90s. Thereafter, my
principal research interests have focused primarily on applying
the latter to the finitary topological, as well as the causet,
scenaria for Lorentzian QG. Thus, after those initial interactions
with Rafael, the resulting scattering saw us taking slightly
different paths towards QG. However, no matter what the future
brings, no matter how much our (re)search (ad)ventures may seem to
differ or diverge from each other, Rafael's paradigmatic figure as
a research scientist and exemplary manner as a human being in
general---his calm, low-key demeanor and mild tone of voice; his
giving you the feeling that he is listening to you quietly, but
attentively and thoughtfully; his impressively deep understanding
and broad knowledge of all the different approaches to QG (and
there is a wild zoo out there!); his original, uncompromising and
`iconoclastic' causet research programme;\footnote{Which, lately,
has been growing from strength to strength, both from gathering
significant results and from gaining popularity.} his kind,
friendly, yet intense, almost ascetic, face, as well as his polite
and inviting manner---will always be with me to inspire and guide
my quests. All in all, I consider myself extremely privileged and
fortunate to have met Rafael personally, and to have engaged, even
if just for a short while, into deep inelastic scattering with him
about QG matters; albeit, well above Planck length(!)

So, {\em belated happy 60th birthday, Rafael}: may you keep on
showing us the way to QG for many years to come, in spite of the
numerous `forks in the road' \cite{sork2}, or of the Sirens' song
of other currently more fashionable QG research programmes, that
may ultimately (mis)lead us astray.

\end{document}